\documentclass[11pt]{elsart}

\usepackage{bbm}
\usepackage{soul}
\usepackage{cite}
\usepackage{graphics,color,epsfig,amsmath,amssymb,bm}
\usepackage{feynmp}
\usepackage{subfigure}
\usepackage[english=british]{csquotes}

\newcounter{bla}

\renewcommand{\half}{\ensuremath{\frac{1}{2}}}

\newcommand{\T}{^{T}}

\newcommand{\be}{\begin{equation}}
\newcommand{\ee}{\end{equation}}
\newcommand{\beqna}{\begin{eqnarray}}
\newcommand{\eeqna}{\end{eqnarray}}

\newcommand{\expectation}{\mathbb{E}}
\newcommand{\w}{\omega}
\newcommand{\wi}{\omega_i}
\newcommand{\wj}{\omega_j}
\newcommand{\wk}{\omega_k}
\newcommand{\wl}{\omega_l}
\newcommand{\wm}{\omega_m}
\newcommand{\wn}{\omega_n}
\renewcommand{\wp}{\omega_p}
\newcommand{\wq}{\omega_q}
\renewcommand{\wr}{\omega_r}
\newcommand{\ws}{\omega_s}

\newcommand{\nOverTwo}{\frac{n}{2}}

\allowdisplaybreaks

\begin{document}

\begin{frontmatter}
\hfill{Report Number: Not Applicable} 

\title{A Probabilistic Angle on One Loop Scalar Integrals}

\author[a]{Kamel Benhaddou}

\address[a]{Unaffiliated \\ almo.cadessi@gmail.com}

\begin{abstract}
Recasting the $N$-point one loop scalar integral as a probabilistic problem, allows the derivation of integral recurrence relations as well as exact analytical expressions in the most common cases.
$\epsilon$ expansions are derived by writing a formula that relates an $N$-point function in decimal dimension to an  $N$-point function in integer dimension. As an example, we give relations for the massive 5-point function in dimension $n=4-2\epsilon$, $n=6-2\epsilon$. The reduction of tensor integrals of rank 2 with $N=5$ is achieved showing the method's potential. Hypergeometric functions are not needed but only integration of arcsine function whose analytical continuation is well known.

\begin{flushleft}
PACS: Not Applicable.
\end{flushleft}


\end{abstract}

\end{frontmatter}

\newpage


\section{Introduction}
There exists a large literature on the problem of $N$-point one loop scalar 
integrals. 
The challenge is to compute $N$-point function for large $N$ in 
dimension $n=4-2\epsilon$ but also for $n>4$ as planned experiment will probe processes like 
$e^{+}\,e^{−}\to → b f_1 \bar{f}^{'}_{1} \bar{b} f_2 \bar{f}^{'}_2$. The so-called 
non-factorisable corrections to $e^{+} e^{-} \to 6f$ require 5-,6-, and 7-point functions ~\cite{staron2006}. In \cite{Davydychev1992}, an application of Mellin Barnes lemma allowed a general series representation for the $N$-point function in terms of Lauricella hypergeometric function $F^D_{N}$ or its generalised version. In \cite{SuzukiSantosSchmidt2003}, it is shown that the negative dimension approach, Mellin Barnes representation and Feynman parameters yield the same expression for the $N$-point function. \cite{Fleischer2003} used a recurrence relation between $N$-point functions in  different dimensions to give an explicit representation of 
the  2-,3-, and 4-point functions in general dimensions $n$. 
Specifically the 4-point function (or box diagram) has received a lot of 
attention. ~\cite{THOOFT1979,vanOldenborgh1990,DENNER1991,VANOLDENBORGH1992,DennerDittmaier2011} perform integration of the Feynman parameters to produce results in terms of a varying number of polylogarithms with real or complex values of the masses.~\cite{vanOldenborgh1990,VANOLDENBORGH1992} shows how to find expressions that are also numerically stable for computer implementation. The 5-point function has also been studied extensively.~\cite{Halpern1963,Melrose1965,Petersson1965,VANNEERVEN1984,BERN1993} 
studied decomposition formulae for the N-point function ($N\geq 5$) in terms of 4-point functions.~\cite{Halpern1963} gave an explicit decomposition for the 5-point function in 4$D$ in terms of 4-point functions in 4$D$, while the others ~\cite{Melrose1965,Petersson1965,VANNEERVEN1984,BERN1993} gave more general 
decompositions formulae.~\cite{Kniehl2010,Kozlov2016} are more specific studies of the massless 5-point function. Dimensional recurrence  proposed in ~\cite{Tarasov1996,Tarasov2000}  is used in ~\cite{Kniehl2010} to obtain an analytic result for the massless pentagon diagram in general $D$ dimensions.~\cite{Kozlov2016} applies the differential equation technique to get a one dimensional integral representation and also discusses $\epsilon$ expansion. In \cite{DavydychevDelbourgo1998} the problem is written as a problem of hyperbolic geometry involving the computation of the volume of a simplex in $D$ dimension. This is obtained when the $N$-point function is cast into the problem of finding the volume of the positive part of an ellipsoid (the volume for which the coordinates are positive). The same problem can actually  be understood probabilistically and describe the probability that a Gaussian random vector has all its components positive. In the following, we rewrite the $N$-point function to make this probabilistic interpretation explicit. It allows us to derive a series expansion in terms of multivariate Hermite polynomials of order $N$ unlike \cite{Davydychev1992} where there are $N+ N(N-1)/2$ summation variables. Unfortunately the Hermite series summation does not appear to be easily resumed in the general case because the explicit form of multivariate Hermite polynomials is still a challenge. Introducing Gaussian random variables, the $N$-point function can be written as an expectation of a Lauricella hypergeometric function $F^D_{N}$ whose arguments depend on the Gaussian variables. Exploiting relations between $F^D_{N}$ and $F^D_{N-1}$, it is possible to write a relation between $N$-point functions e.g. an integral relation linking the $N$-point function in dimension $n$ to the $(N-1)$-point function in dimension $n-2$. In ~\cite{DavydychevDelbourgo1998}, the geometrical interpretation was pushed as far as $N=4$. Here, we show that our methods can easily handle cases of $N>4$. We give explicit computations for $N=2,3,4,5,6,7$ in integer dimension. No hypergeometric functions are required to express the results which are given by integrals of arcsine functions. When the dimension $n$ is smaller than $N$, the $N$-point function can be given by a large number of terms as is seen for the 6-point function in dimension 4. An $N$-point function is also characterised by the powers $\nu_i$ of each propagator. When $\nu=\sum_{i}\nu_i$  is greater (smaller) than the spacetime dimension $n$, the $N$-point function in dimension $n$ can be obtained by integrating in the complex plane (real positive line) an $N$-point function in dimension $N$. This integral representation is a compact way of expressing $N$-point function in dimension $n \neq N$. All order $\epsilon$ expansions can be obtained from this integral representation as the dimension $n$ is decoupled from the other variables in the integrand. Starting from the more familiar Feynman representation, $\epsilon$ expansion is obtained as an integral relating a general $N$-point function in $d-2\epsilon$ dimension to the same $N$-point function with an integer dimension $d-2k, k\in \mathbf{N}$ and modified mass parameters. Each term of the epsilon expansion is an integral of an $N$-point function in integer dimension. Having shown how to perform $\epsilon$ expansion as well as how to compute diagrams with power $\nu_i>1$, allows us, using Davydychev formula ~\cite{Davydychev1991}, a complete reduction of tensor integrals. This is shown explicitly for the tensor integral of rank $r=2$ with $N=5$. The extension of the reduction program to $N>5$ and $r>2$ is left for the future.

\section{Probability Theory} 
In this section, we summarise some results from probability theory ~\cite{Sinai2012,Venkatesh2012}. This will set some notations as well as help those less familiar with the concepts used in probability theory. 

We can consider a random variable as a function that associates a numerical value to a given outcome of an experiment. If the set of possible outcomes if finite (e.g. as in throwing a dice), the random variable is said to be discrete. When the set of possible outcome is continuous, the random variable is said to be continuous.

\subsection{Density and Distribution Functions}
To study continuous random variables, the concept of probability density function is introduced.
The probability density function $f_X(x)$ gives the probability that the random variable $X$ will fall within a certain interval. We have
\be\label{density_def}
\mathrm{Prob}( x < X < x+\mathrm{d}x) = f_X(x) \mathrm{d}x\,. 
\ee
Let us consider a continuous random variable $X$ for which the set of possible outcome is $\mathbb{R}$. Because any outcome should be in $\mathbb{R}$, the probability density function is normalised as follows
\be
\int_{-\infty}^{\infty} f_X(x) \, \mathrm{d}x=1\,.
\ee
 By integrating Eq.(\ref{density_def}), we can find the probability that the random variable  $X$ will fall in a given interval $[u,v]$
\be
\mathrm{Prob}( u < X < v) = \int_{u}^{v}\ f_X(x) \mathrm{d}x\,. 
\ee 
The \enquote{cumulative distribution function} of $X$, $F_X(x)$ gives the probability that the random variable will be smaller than a threshold $x$. It is defined as follows
\be
F_X(x) = \int_{-\infty}^x f_X(u) \, du\,,
\ee
so the density function is obtained from the distribution function by differentiation
\be
f_X(x) = \frac{\mathrm{d}}{\mathrm{d}x} F_X(x)\,.
\ee
For continuous random variables $X_1,X_2,\ldots,X_n$, we define the \enquote{joint probability density function} $f_{X_1,\ldots,X_n}\left(x_1,x_2,\ldots,x_n\right)$. For a domain
 $\mathbb{D}\in\mathbb{R}$, the probability that each random variable $X_i, i=1,\ldots,n$ falls in the domain $\mathbb{D}$ is
\be
\mathrm{Prob} \left[ \left(X_1,X_2,\ldots,X_n\right)\in \mathbb{D}\right]
= \int\limits_{-\infty}^{\infty}\mathrm{d}x_1
\ldots
\int\limits_{-\infty}^{\infty}\mathrm{d}x_n
\prod_{i=1}^{n} \mathbbm{1}_{x_i\in \mathbb{D}}\, f_{X_1,\ldots,X_n}\left(x_1,x_2,\ldots,x_n\right)\,.
\ee
To simplify the notation, we introduce the following vector notation
\begin{align}
X &=  \left(X_1,X_2,\ldots,X_n\right)\,,\nonumber \\
x &=  \left(x_1,x_2,\ldots,x_n\right)\,,\nonumber \\
f_X(x) &= f_{X_1,\ldots,X_n}\left(x_1,x_2,\ldots,x_n\right) \,.
\end{align}
It should be clear from the context whether we are working with unit or multivariate densities.
The most ubiquitous random variable is the so called normal (or Gaussian) random variable whose density function $f_X^{G}(x)$is given by
\be
f_X^{G}(x) = \frac{1}{\sqrt{2\pi \sigma^2}} e^{-\half\frac{\left( x- \mu\right)^2}{\sigma^2}}\,,
\ee
where $\mu \in \mathbb{R}$ and $\sigma>0$ are two parameters whose meaning will be given later.  

\subsection{Expected Value, Moments, Variance and Covariance}
The expected value of a random variable $X$, can be intuitively understood as the average value of the outcomes after a large number of experiments has been performed. Given the density 
function $f_X(x)$ the expected value $\expectation[X]$ is computed as follows
\be  
\expectation[X] = \int_{-\infty}^{\infty} x f_{X}(x) \,\mathrm{d}x\,.
\ee
Given a functional form $g(X)$, which can represent some quantity one is interested in, the expected value $\expectation[g(X)]$ of $g(X)$ is
\be
\expectation\left[ g(X) \right] = \int_{-\infty}^{\infty} g(x) f_{X}(x) \,\mathrm{d}x\,.
\ee

For $g(X) = X^n, n\in \mathbb{N}$, we obtain the raw moments $\mu'_n$
\be
\mu'_n(X) = \int_{-\infty}^{\infty} x^n f_{X}(x) \,\mathrm{d}x\,.
\ee

For $g(X) = \left( X - \expectation[X]\right)^n, n\in \mathbb{N}$, we obtain the central moments $\mu_n$
\be
\mu_n(X) = \int_{-\infty}^{\infty} \left( x - \mu\right)^n f_{X}(x) \,\mathrm{d}x\,.
\ee
with $\mu = \expectation[X]$. The second central moment $\mu_2$ is called the variance of the random variable $X$
\begin{align}
\mu_2 &= \mathrm{Var}(X) \,,\nonumber \\
&= \int_{-\infty}^{\infty} \left( x - \mu\right)^2 f_{X}(x) \,\mathrm{d}x\,.
\end{align}
The variance $\mathrm{Var}(X)$ gives the averaged squared deviation from the expected 
value $\expectation[X]$

Given a joint  probability density function $f_{X_1,\ldots,X_n}(x_1,\ldots,x_n)$ the raw moments 
are denoted $\mu_{k_1,\ldots,k_n}$
\be
\mu_{k_1,\ldots,k_n} = 
\int_{-\infty}^{\infty}\mathrm{d}x_1 \ldots \int_{-\infty}^{\infty}\mathrm{d}x_n
\, x_1^{k_1}\ldots x_n^{k_n} \,
f_{X_1,\ldots,X_n}(x_1,\ldots,x_n)\,.
\ee
Truncated moments $\mu^{tr}_{k_1,\ldots,k_n}$ are obtained when some or all of the integrations are truncated by introducing some thresholds $z_i$ for the $X_i$ variables
\be
\mu^{tr}_{k_1,\ldots,k_n}\left(z_1,\ldots,z_n \right) = 
\int_{-\infty}^{\infty}\mathrm{d}x_1 \ldots \int_{-\infty}^{\infty}\mathrm{d}x_n
\, x_1^{k_1}\mathbbm{1}_{x_1 > z_1}\ldots x_n^{k_n}\mathbbm{1}_{x_n > z_n} \,
f_{X_1,\ldots,X_n}(x_1,\ldots,x_n)\,.
\ee
In general the expectation operator $\expectation$ and functions  
of random variables don't commute that is for a non linear function $g(X)$ we have
\begin{align}
\expectation[g(X)] &= 
\int_{-\infty}^{\infty}\mathrm{d}x \, g(x) f_X(x)\,, \nonumber \\
&  \neq g(\expectation[X])\,.
\end{align}
If the functional form $g(X)$ is given by an integral, we can still commute the integral and the expectation operator $\expectation$ as it corresponds to interchanging the order of integration
\begin{align}
\expectation\left[ \int_{a}^{b} \mathrm{d}u \, g(X, u)  \right] &= 
\int_{-\infty}^{\infty}\mathrm{d}x \, \left [
\int_{a}^{b} \mathrm{d}u\,  g(x,u)\right] f_X(x)\,, \nonumber \\
&= \int_{a}^{b} \mathrm{d}u\,  \left [\int_{-\infty}^{\infty}\mathrm{d}x\, g(x,u) f_X(x)
 \right] \,, \nonumber \\
&= \int_{a}^{b} \mathrm{d}u\,\expectation\left[ g(X,u)\right]\,.
\end{align} 
Two random variables $X$ and $Y$ are said to be independent if the events $X\leq x$ and $Y\leq y$ are independent i.e. one event has no influence on the other and vice versa. For independent random variables $X$ and $Y$, both the distribution $F_{X,Y}(x,y)$ and density functions $f_{X,Y}(x,y)$  satisfy
\begin{align}
F_{X,Y}(x,y) &= F_{X}(x) F_{Y}(y)\,,\nonumber \\
f_{X,Y}(x,y) &= f_{X}(x) f_{Y}(y)\,.\nonumber
\end{align} 
Also, the expected value of the product is the product of the expected values
\begin{align}
\expectation\left[ X Y\right] &= 
\int_{-\infty}^{\infty}\mathrm{d}x
\int_{-\infty}^{\infty}\mathrm{d}y\, 
xy f_{X,Y}(x,y)\,,\nonumber \\
&= 
\int_{-\infty}^{\infty}\mathrm{d}x\, x f_{X}(x) 
\int_{-\infty}^{\infty}\mathrm{d}y\, y f_{Y}(y)\,,\nonumber \\
&= \expectation[X] \expectation[Y]\,.
\end{align}

The covariance $\mathrm{COV}(X,Y)$ between two jointly distributed random variables $X$ and $Y$ is defined 
as
\begin{align}
\mathrm{COV}(X,Y) &= \expectation\left[\left(X -\expectation[X]\right) \left(Y -\expectation[Y]\right)\right]\,, \nonumber \\
&= \expectation[XY] - \expectation[X]\expectation[Y]\,, \nonumber \\
&= \mathrm{COV}(Y,X)\,.
\end{align}
The covariance measures how much changes in one random variable affect the other. When $X$ 
and  $Y$ are independent the covariance is null. 
The covariance between $X$ and itself is the variance
\be
\mathrm{COV}(X,X) =  \mathrm{Var}(X)\,.
\ee
From the covariance, we define the linear correlation coefficient $\rho_{XY}$ 
as
\begin{align}
\rho_{XY} &= \frac{\mathrm{COV}(X,Y)}{\sqrt{\mathrm{Var}(X) \mathrm{Var}(Y)}}\,, \nonumber \\
&=\rho_{YX}
\end{align}
Let $X$ and $Y$ be random variables with non-zero variances. Then the absolute value of the correlation coefficient $\rho_{XY}$ is less than or equal to one. If $|\rho_{XY}| = 1$, then for some constants a and b the equality $Y = a X + b$ holds almost surely.

Let $X = (X_1,\ldots,X_n)$ be a random vector. The covariance matrix $\Sigma$ is a matrix whose 
$(i,j)$ component is the covariance between $X_i$ and $X_j$. The components of the matrix 
($i\neq j$) are
\begin{align}
\Sigma_{ii} &= \mathrm{COV}(X_i,X_i) = \mathrm{Var}(X_i)\,, \nonumber \\
\Sigma_{ij} &= \mathrm{COV}(X_i,X_j) \,, \nonumber \\
            &= \sqrt{\mathrm{Var}(X_i)} \sqrt{\mathrm{Var}(X_j)} \rho_{X_i,X_j}\,.
\end{align}

\subsection{Characteristic Functions and Multivariate Normal Distribution}
The characteristic function $\phi_X(s)$, with $s\in \mathbb{R}$,  of a random variable $X$ is given by the expected value of $e^{-i s X}$ 
\begin{align}
\phi_X(s) &= \expectation\left[ e^{-i s X} \right]\,, \nonumber \\
&= \int_{-\infty}^{\infty}\mathrm{d}x\, e^{-i s x} f_X(x)\,.
\end{align}
The characteristic function of a normal random variable with parameter $\mu$ and $\sigma^2$ is 
\be
\phi_{\mu,\sigma^2}(s) = e^{-\half \sigma^2 s^2 -i \mu s}\,.
\ee
Given a vector $s=(s_1,\ldots,s_n)$, we define the multivariate characteristic function as the 
following expectation value
\be
\phi_{X_1,\ldots,X_n}(s_1,\ldots,s_n) = \expectation\left[ e^{-i (s_1X_1 +\ldots+s_n X_n)}\right]\,.
\ee
Introducing the vector $X=(X_1,\ldots,X_n)$, we can rewrite it as
\be
\phi_{X_1,\ldots,X_n}(s) = \expectation\left[ e^{-i (s^T.X)}\right]\,,
\ee 
where $T$ is the transpose operator and the dot is the matrix multiplication.

A random vector $X = (X_1,\ldots, X_n)$ is said to have the multivariate normal distribution if there is a $n$-vector $\mu$ and a symmetric, positive semi definite 
$n\times n$ matrix $\Sigma$, such that the characteristic function of $X$ is
\be\label{characteristic_function_multinormal}
\phi_{X}(s) = e^{-\half s^T . \Sigma . s - i \mu\T.s}\,.
\ee 
The joint density of the $X_i$ is 
\be
f_{X_1,\ldots,X_n} (x_1,\ldots,x_n) = \frac{1}{(2 \pi)^{\frac{n}{2}}|\Sigma|^{\half}}
e^{-\half \left( x - \mu\right)^T . \Sigma^{-1}. \left( x - \mu\right)}\,,
\ee
where $|\Sigma|$ is the determinant of $\Sigma$.

The orthant probabilities $P_0(\Sigma)$ is the probability that a multivariate normal 
random vector with zero mean has all his components positive (equivalently negative)
\begin{align}
P_0(\Sigma)  & =  \expectation\left[ \prod_{i=1}^{n}\mathbbm{1}_{X_i>0}\right]\,, \nonumber\\
&=
\frac{1}{(2 \pi)^{\frac{n}{2}}|\Sigma|^{\half}}
\int\limits_{0}^{\infty}\mathrm{d}x_1
\ldots
\int\limits_{0}^{\infty}\mathrm{d}x_n
\,
e^{-\half x\T .\Sigma^{-1}. x}\,.
\end{align}
It will be seen in the next sections that the$N$-point function in dimension $n=N$ can be interpreted as an orthant probability. The matrix $\Sigma$ has components that depends on the kinematical variables of the $N$-point diagram.

\section{N Point One Loop Scalar Integral}
We  consider the $N$-point one loop scalar integral in 
$n$ dimensions and with each propagator having power $\nu_i$ 
\be
J^{N}(n;\nu_1, \nu_2, \ldots, \nu_N) = \int\,\frac{\mathrm{d}^nq}{i\pi^{\frac{n}{2}}}
\frac{1}{\prod_{i=1}^{N}\left[ (p_i + q)^2 - m_i^2\right]^{\nu_i}}\,
\ee
which after introduction of Feynman parameters takes the following form ~\cite{DavydychevDelbourgo1998}
\begin{align}
\label{JN_feynman}J^{N}(n;\bm{\nu};\Sigma) &= \frac{1}{(-1)^{\nu}} \frac{\Gamma(\nu - n/2)}{\prod_i \Gamma(\nu_i)}
\int_{0}^{1}\mathrm{d}u_1 u_1^{\nu_1-1} \ldots \int_{0}^{1}\mathrm{d}u_N u_N^{\nu_N-1}   
\frac{\delta\left( \sum\limits_{i=1}^{N} u_i -1\right)}{\left( -\sum\limits_{j<\ell}u_j u_{\ell} k^2_{j\ell} + \sum\limits_i u_i m_i^2   \right)^{\nu - \frac{n}{2}}}\,, \\
\label{JN_matrix}&= (-1)^{-\nu} \frac{\Gamma(\nu - n/2)}{\prod_i \Gamma(\nu_i)}
\int_{0}^{1}\mathrm{d}u_1 u_1^{\nu_1-1} \ldots \int_{0}^{1}\mathrm{d}u_N u_N^{\nu_N-1}   
\frac{\delta\left( \sum\limits_{i=1}^{N} u_i -1\right)}{\left(u\T . \Sigma .u \right)^{\nu - \frac{n}{2}}}\,,
\end{align}
where $(\sum\limits_i \nu_i =\nu$) and the matrix $\Sigma$ has components
\begin{align}
\Sigma_{ii} &= m_i^2\,, \nonumber \\
\Sigma_{j\ell} &= m_j m_\ell c_{j\ell}   & j\neq \ell\,, \nonumber \\
c_{j\ell} &= \frac{m_j^2 + m_{\ell}^2 - k_{j\ell}^2}{2 m_j m_{\ell}}\,, \nonumber \\
k_{j\ell}^2 &= \left( p_j - p_{\ell}\right)^2\,. \nonumber
\end{align}
The $c_{j\ell}$ can be understood as cosines of some angles $\tau_{j\ell}$ ~\cite{DavydychevDelbourgo1998}
\be\label{def_costau}
c_{j\ell} = \cos\tau_{j\ell} 
= \left\{ \begin{array}{c} \;\; 1, \;\;\; k_{j\ell}^2=(m_j-m_{\ell})^2 \\
                               -1, \;\;\; k_{j\ell}^2=(m_j+m_{\ell})^2 
          \end{array} \right. \; .
\ee
The corresponding angles $\tau_{j\ell}$ are ~\cite{DavydychevDelbourgo1998}
\be\label{def_tau}
\tau_{j\ell}= \arccos(c_{j\ell}) 
=\arccos\left(\frac{m_j^2+m_{\ell}^2-k_{jl}^2}{2m_j m_{\ell}}\right)
= \left\{ \begin{array}{c}   0, \;\;\; k_{jl}^2=(m_j-m_{\ell})^2 \\
                           \pi, \;\;\; k_{jl}^2=(m_j+m_{\ell})^2
          \end{array} \right. \; .
\ee
The angles $\tau_{j\ell}$ can be analytically continued when the $c_{j\ell}$ are not 
in the range $[-1,1]$. When $k_{j\ell}^2<(m_j-m_l)^2$,  the $c_{jl}$ are greater than 1 and the angles $\tau_{j\ell}$ are given by ~\cite{DavydychevDelbourgo1998}
\begin{align}
\tau_{j\ell} &=-\mathrm{i}\; \mathrm{Arch}(c_{j\ell})\,,\\
&= -\frac{\mathrm{i}}{2} 
\ln\left( 
\frac{m_j^2+m_{\ell}^2-k_{j\ell}^2+\sqrt{\lambda(m_j^2,m_{\ell}^2,k_{j\ell}^2)}}
     {m_j^2+m_{\ell}^2-k_{j\ell}^2-\sqrt{\lambda(m_j^2,m_{\ell}^2,k_{j\ell}^2)}}
\right) ,
\end{align}
where $\lambda(x,y,z)$ is the K\"{a}llen function
\be
\lambda(x,y,z) = x^2 + y^2 + z^2 -2xy -2xz -2yz\,.
\ee
When $k_{j\ell}^2>(m_j+m_{\ell})^2$, the $c_{j\ell}$ are smaller than 1 and the angles $\tau_{j\ell}$ are given by ~\cite{DavydychevDelbourgo1998}
\begin{align}
\tau_{j\ell} &=\pi+\mathrm{i}\; \mathrm{Arch}(-c_{j\ell})\,, \\
&= \pi + \frac{\mathrm{i}}{2}
\ln\left(
\frac{k_{jl}^2-m_j^2-m_{\ell}^2+\sqrt{\lambda(m_j^2,m_{\ell}^2,k_{j\ell}^2)}}
     {k_{jl}^2-m_j^2-m_{\ell}^2-\sqrt{\lambda(m_j^2,m_{\ell}^2,k_{j\ell}^2)}}
\right)\,.
\end{align}
When the angles  $\tau_{j\ell}$ are real (as in Eq.(\ref{def_tau})), the $c_{j\ell}$ are in the range $[-1,1]$ and the matrix $\Sigma$ in Eq.(\ref{JN_matrix}) can be interpreted as a covariance matrix. Let us consider a random vector $Z$ of size $N$, whose components are normal random variables $Z_i$ with 0 mean and variance $\sigma^2_i = m_i^2>0$. When the $c_{j\ell}$ are in the range $[-1,1]$, we can interpret the $c_{j\ell}$ as the correlation between $Z_j$ and $Z_{\ell}$.   

Using the relation
\be\label{Schwinger_parameter}
\frac{\Gamma(\alpha)}{A^\alpha} = \int_{0}^{\infty}\mathrm{d} \tau\, \tau^{\alpha-1} e^{-\tau A}\,,
\ee
Eq.(\ref{JN_matrix}) is rewritten as
\be
J^{N}(n;\bm{\nu};\Sigma) =  \frac{(-1)^{-\nu}}{\prod_i \Gamma(\nu_i)}
\int_{0}^{1}\mathrm{d}u_1 u_1^{\nu_1-1} \ldots \int_{0}^{1}\mathrm{d}u_N u_N^{\nu_N-1}   
\delta\left( \sum\limits_{i=1}^{N} u_i -1\right)
\int_{0}^{\infty}\, \mathrm{d}\tau\,  \tau^{\nu -\frac{n}{2}-1} e^{-\tau u\T .\Sigma. u }\,.
\ee
We change the variable $\tau$ to $v$ such that $\tau = \frac{v^2}{2}$ and obtain
\be\label{J_N_v_integration}
J^{N}(n;\bm{\nu};\Sigma) = (-1)^{-\nu}
\int_{0}^{1}\mathrm{d}u_1 u_1^{\nu_1-1} \ldots \int_{0}^{1}\mathrm{d}u_N u_N^{\nu_N-1}   
\frac{\delta\left( \sum\limits_{i=1}^{N} u_i -1\right)}
{2^{\nu -\nOverTwo-1}\prod_i \Gamma(\nu_i)} 
\int_{0}^{\infty}\, \mathrm{d}v\, v^{2\nu -n-1} e^{-\half (vu)\T .\Sigma. (vu) }\,.
\ee
Let $Z = (Z_1,\ldots,Z_N)$ be a $N$ dimensional random vector distributed according to the multivariate normal distribution with zero mean vector and covariance matrix $C$, then its characteristic function 
$\phi(s_1,\ldots,s_n) = \phi(s)$ is
\be
\phi(s) = \expectation  \left[ e^{-i s\T .Z }\right] = e^{-\half s\T .C. s}\,.
\ee
Choosing an $N$ dimensional random vector $Z$ distributed normally with zero mean and  covariance matrix $\Sigma$, we can linearise the quadratic form $(vu)\T .\Sigma. (vu)$ appearing in the exponential and perform the integration over $v$ using Eq.(\ref{Schwinger_parameter})
\begin{align}
J^{N}(n;\bm{\nu};\Sigma) &=  (-1)^{-\nu}
\int_{0}^{1}\mathrm{d}u_1 u_1^{\nu_1-1} \ldots \int_{0}^{1}\mathrm{d}u_N u_N^{\nu_N-1}   
\frac{\delta\left( \sum\limits_{i=1}^{N} u_i -1\right)}
{2^{\nu -\nOverTwo-1}\prod_i \Gamma(\nu_i)} 
\int_{0}^{\infty}\, \mathrm{d}v\, v^{2\nu -n-1} \expectation  \left[e^{-i v u\T .Z} \right]\,, \nonumber \\
&  \nonumber \\
\label{JN_Feynman_rep}&= \frac{(-1)^{-\nu}\Gamma(2\nu -n)}{2^{\nu -\nOverTwo-1}\prod_i \Gamma(\nu_i)}
\expectation  \left[
\int_{0}^{1}\mathrm{d}u_1 u_1^{\nu_1-1} \ldots \int_{0}^{1}\mathrm{d}u_N u_N^{\nu_N-1}   
\frac{\delta\left( \sum\limits_{i=1}^{N} u_i -1\right)}{\left( i u\T .Z\right)^{2\nu -n}}\right]\,.  
\end{align}
Let us write 
\be
i u\T .Z = i \sum\limits_{j=1}^{N} u_j Z_j = 1 + \sum\limits_{j=1}^{N} u_j (-1+i Z_j)\, \nonumber
\ee
and use the Mellin Barnes 
representation ~\cite[Eq.(3.4)]{Davydychev1991_2}
\beqna
\frac{\Gamma(c)}{\left(A_1+A_2+\ldots + A_n\right)^{c}} 
&=&\int\limits_{-i\infty}^{i\infty} \frac{\mathrm{d}s_1}{2\pi i} \ldots \int\limits_{-i\infty}^{i\infty} \frac{\mathrm{d}s_{n-1}}{2\pi i} 
\Gamma(-s_1)\ldots \Gamma(-s_{n-1})  \Gamma(c + s_1 + \ldots +s_{n-1})\,, \nonumber \\
&& A_1^{s_1} A_2^{s_2} \ldots A_{n-1}^{s_{n-1}} A_{n}^{-c -s_1 - \ldots - s_{n-1}}\,,
\eeqna
to get
\begin{align}
J^{N}(n;\bm{\nu};\Sigma) &=
\expectation  \left[ (-1)^{-\nu}
\int_{0}^{1}\mathrm{d}u_1 u_1^{\nu_1-1} \ldots \int_{0}^{1}\mathrm{d}u_N u_N^{\nu_N-1}   
\frac{\delta\left( \sum\limits_{i=1}^{N} u_i -1\right)} 
{2^{\nu -\nOverTwo-1}\prod_i \Gamma(\nu_i)} \right.\nonumber \\
& \int\limits_{-i\infty}^{i\infty} \frac{\mathrm{d}s_1}{2\pi i} \ldots \int\limits_{-i\infty}^{i\infty} \frac{\mathrm{d}s_N}{2\pi i} 
\Gamma(-s_1) \ldots \Gamma(-s_N)  \Gamma(2 \nu -n + \sum\limits_{i=1}^{N} s_i)   \nonumber \\
& \left. \,  \prod_{j=1}^{N} \left( u_j(i Z_j -1)\right)^{s_j}\right]\,.  \nonumber
\end{align}
The integration over the $u$ variables is analytical as well as the integration in the complex plane as the 
$\Gamma(s)$ function has poles at $s=- m$ with $m$ integer and residue $(-1)^m/m!$. So 
\be\label{JN_series}
J^{N}(n;\bm{\nu};\Sigma) =  \frac{(-1)^{-\nu}}{2^{\nu -\nOverTwo-1}\prod_i \Gamma(\nu_i)}
\expectation  \left[ 
\sum\limits_{k_1=0}^{\infty}\ldots\sum\limits_{k_N=0}^{\infty}\, \frac{\Gamma(2\nu -n + \sum\limits_{j} k_j)}{\Gamma(\nu + \sum\limits_j k_j)} \prod_{j=1}^{N} \Gamma(\nu_j + k_j) \prod_{j=1}^{N} \frac{(1-i Z_j)^{k_j}}{k_j!}
\right]\,,
\ee 
which can be written as a Lauricella $F_D$ hypergeometric function
\be\label{JN_lauricella_expectation}
J^{N}(n;\bm{\nu};\Sigma) =  \frac{(-1)^{-\nu}}{2^{\nu -\nOverTwo-1}} 
\frac{\Gamma(2\nu -n )}{\Gamma(\nu)} 
\expectation  \left[ 
F_D\left(2\nu -n; \nu_1, \ldots, \nu_N; \nu; 1-iZ_1,\ldots, 1-iZ_N\right)
\right]\,.
\ee
The Lauricella hypergeometric function $F_D$ can be expressed as a series
\begin{align}
F_D(a; b_1,b_2,\ldots,b_N;c; x_1,x_2,\ldots,x_N) &= \sum\limits_{k_1=0}^{\infty}\ldots\sum\limits_{k_N=0}^{\infty}\, \frac{(a)_{\sum k_i}}{(c)_{\sum k_i}}\prod_{i=1}^{N}(b_i)_{k_i} \prod_{i=1}^{N}\frac{x_i^{k_i}}{k_i!}\,, \nonumber \\
& |x_i| < 1\,, i=1,\ldots,n\,,
\end{align}
and it is this form that first occurred in the computation of N-point function ~\cite[Eq.(2.9)]{Davydychev1991_2}. It also has an integral representation (that can be easily derived from the series representation)
\begin{align}
F_D(a; b_1,b_2,\ldots,b_N;c; x_1,x_2,\ldots,x_N) &=
\frac{\Gamma(c)}{\Gamma(a)\Gamma(c-a)}\int_{0}^{1}\mathrm{d}t \,t^{a-1} (1-t)^{c-a-1}
\frac{1}{\prod_{i=1}^{N} \left(1 - x_i t\right)^{b_i}}\,,\nonumber \\
&\label{FD_integral_representation} \Re(c) > \Re(a)>0\,.
\end{align}
In Eq.(\ref{JN_series}), if we take the expectation over the normally distributed random variables, we obtain a series in term of multivariate Hermite polynomials. In ~\cite{Withers2000}, it was shown that the multivariate Hermite polynomials $H_{\mathrm{e}k_1k_2 \ldots k_p}(x, \Sigma)$ in $p$ dimensions have the probabilistic representation
\be
H_{\mathrm{e}k_1k_2 \ldots k_p}(x, \Sigma) = \expectation  \left[ \prod_{j=1}^{p} 
\left( (\Sigma^{-1}.x)_j + i Z_j\right)^{k_j}\right]\,,
\ee
where  $Z$ is a $p$ dimensional random vector normally distributed with zero mean and covariance matrix $\Sigma^{-1}$. We have
\be\label{JN_series_hermite}
J^{N}(n;\bm{\nu};\Sigma) =  \frac{(-1)^{-\nu}}{2^{\nu -\nOverTwo-1}\prod_i \Gamma(\nu_i)} 
\sum\limits_{k=0}^{\infty}\, \frac{\Gamma(2\nu -n + k)}{\Gamma(\nu + k)} 
\sum\limits_{\sum k_i=k}  \frac{\Gamma(\nu_i + k_i)}{k_i!}H_{\mathrm{e}k_1\ldots k_N}(\Sigma.\mathbf{1},\Sigma)\,,
\ee  
where $\mathbf{1}$ is a vector with all components equal to 1. This is a multi sum with $N$ indices to be contrasted with the series obtained in ~\cite{Davydychev1992} which involves $N^2$ indices. As the computation of multivariate Hermite polynomial is still a challenge, this series seems at first of limited numerical interest but it might be useful to obtain asymptotic expansions by considering the first few terms ($k=0,1,2$) of the series. The Hermite polynomials 
are given by expectation of products of normal random variables, which can be computed by application of the Wick Theorem, as in done in path integral perturbation theory for connected $N$-point functions.

In the following we exploit the form given by the expectation of the Lauricella hypergeometric function $F_D$  with parameters
\begin{align}
a &= 2\nu -n\,, \nonumber \\
b_i &= \nu_i\,,  \nonumber \\
c&=  \nu\,. \nonumber
\end{align}
We distinguish three cases

(\bm{$\nu -n <0.$}) 
In this case $c>a$ so we use the one dimensional integral 
representation Eq.(\ref{FD_integral_representation}) to write 
\be
J^{N}(n;\bm{\nu};\Sigma) =  \frac{(-1)^{-\nu}}{2^{\nu -\nOverTwo-1}\Gamma(n-\nu)}
\expectation  \left[
\int_{0}^{1}\mathrm{d}t \,t^{2\nu - n-1} (1-t)^{n-\nu-1}
\frac{1}{\prod_{i=1}^{N} \left(1 - \theta_i t\right)^{\nu_i}}\right]\,,
\ee
with $\theta_i = 1 - i Z_i$.

(\bm{$\nu = n.$})
In this case, the two gamma functions in Eq.(\ref{JN_series}) disappear and using
\be
\frac{\Gamma(\nu)}{(1-x)^\nu} = \sum\limits_{k=0}^{\infty}\Gamma(\nu+k)\frac{x^k}{k!}\,,
\ee 
we obtain
\be
J^{N}(n;\bm{\nu};\Sigma) =  \frac{(-1)^{-\nu}}{2^{\nu -\nOverTwo-1}}
\expectation  \left[  
\prod_{i=1}^{N} \frac{1}{(i Z_{i})^{\nu_i}} \right]\,.
\ee

(\bm{$\nu - n = k >0.$})
In this case we go back to the Feynman representation Eq.(\ref{JN_Feynman_rep})
\be
J^{N}(n;\bm{\nu};\Sigma)=
 \frac{(-1)^{-\nu}\Gamma(2\nu -n)}{2^{\nu -\nOverTwo-1}\prod_i \Gamma(\nu_i)}
\expectation  \left[ \int \prod_i \mathrm{d}u_i u_i^{\nu_i-1} \delta\left( \sum u_i -1\right)
\frac{1}{\left( i u\T .Z\right)^{2\nu -n}}\right]\,.
\ee
We insert in the numerator
\be
1 = \left(\sum\limits_{i=1}^{N} u_i\right)^k = \sum\limits_{\sum k_i=k}\, \frac{k!}{\prod_i k_i!}\prod_i u_i^{k_i}\,,
\ee
and obtain
\begin{align}
J^{N}(n;\bm{\nu};\Sigma)&=
 \frac{(-1)^{-\nu}}{2^{\nu -\nOverTwo-1}\prod_i\Gamma(\nu_i)}
\expectation  \left[ \sum\limits_{\sum k_i = k} \frac{k! \prod_i \Gamma(\nu_i +k_i)}{\prod_i k_i!}
\, \prod_i \frac{1}{(iZ_i)^{\nu_i+k_i}}\right]\,, \\
&=\label{JN_k_greater_than_zero}(-1)^{k}
\sum\limits_{\sum k_i = k} \frac{k! \prod_i \left(\nu_i\right)_{k_i}}{\prod_i k_i!}
\, J^{N}(\nu+ k; \bm{\nu}+\bm{k}; \Sigma)\, 
\end{align}
where $\bm{k}$ is a vector whose $i$th component is $k_i$ and $(a)_x$ is the Pochhammer symbol
\be
(a)_x = \frac{\Gamma(a+x)}{\Gamma(a)}\,.
\ee 
If we apply this for $k=1$, we find 
\be
J^{N}(\nu - 1;\bm{\nu};\Sigma)= - \sum\limits_{j=1}^{N} \nu_j J^{N}(\nu+ 1; \bm{\nu}+\bm{\delta_{j}}; \Sigma)\,,
\ee
where $\bm{\delta_{j}}$ indicates a vector whose $j$th component is 1. This particular case is also a particular case of a formula derived in ~\cite{Davydychev1991} and 
later in ~\cite{Tarasov1996}.   

\section{Relations Between $N$-Point Functions}
By exploiting the different representation of the $F_D$ function it is possible to derive relations between the $N$-point functions. The parameter $c$ of the $F_D$ function is equal to the sum of the $b_i$ parameters, so it can be written as the so called Carlson $R$ function ~\cite{Carlson1963}
\begin{align}
R(a;\bm{b};\bm{z})  &= R(a;b_1,b_2,\ldots,b_N;z_1,z_2,\ldots,z_N)\,, \\ 
&= F_D(a;b_1,b_2,\ldots,b_N;b_1+b_2+\ldots+b_N;1-z_1,1-z_2,\ldots,1-z_N)\,.\nonumber
\end{align}
so
\begin{align}
J^{N}(n;\bm{\nu};\Sigma) &=  \frac{(-1)^{-\nu}}{2^{\nu -\nOverTwo-1}} 
\frac{\Gamma(2\nu -n )}{\Gamma(\nu)} 
\expectation  \left[ R\left(2\nu -n; \bm{\nu}; i\bm{Z}\right) \right]\,, \\
&=
\frac{(-1)^{-\nu}}{2^{\nu -\nOverTwo-1}} 
\frac{\Gamma(2\nu -n )}{\Gamma(\nu)}\,\, 
\bar{J}^{N}(n;\bm{\nu};\Sigma)\,,
\end{align}
where we have defined the rescaled $N$-point function $\bar{J}^{N}(n,\bm{\nu}, \Sigma)$ for ease of use.

In ~\cite{Kreimer1992,Kreimer1993,BrucherFranzkowskiKreimer1994}, 
explicit representations for the scalar and tensor 2,3-point functions were given in 
term of the $R$ function. In our representation, we have the $R$ function but we still need to compute the expectation over the Gaussian random variables, which is a non trivial task in the general case. Instead, we are going to exploit relations satisfied by the $R$ function to write relations between the $N$-point functions.
Let us define
\be
B(b_1,b_2,\ldots,b_N) = \frac{\Gamma(b_1)\ldots\Gamma(b_N)}{\Gamma(b_1+b_2+\ldots+b_N)}\,.
\ee

An interesting relation for the function $R$ is an integral representation that gives $R$ as an integral of an $R$ function with one less variable ~\cite[Eq.(7.4)]{Carlson1963}	
\be\label{R_recurrence_1}
B(c-b_N,b_N) R(a;\bm{b}; \bm{z}) = \int_{0}^{\infty}\mathrm{d}t\,\frac{t^{b_N-1}}{(1+t)^{a'}}
R(a;b_1,\ldots,b_{N-1};z_1+tz_N, \ldots, z_{N-1}+t z_{N})\,, 
\ee	
for $\Re(c) > \Re(b_N)	>0$.
Let $\eta_i, i=1,\ldots,N-1$ be normal random variables defined as
\be
\eta_i = Z_i + t Z_N\,.
\ee	
The $\eta_i$ have covariance matrix $\Sigma^{\eta}(t)$ with components
\begin{align}
\Sigma^{\eta}_{j\ell}(t) &= \expectation  \left[ (Z_j + t Z_N)(Z_{\ell} + t Z_N) \right] \,,\,\,\,\,\,\,\,\,\,  1 \leq j,\ell \leq N-1\,,   \\ 
&= \Sigma_{j\ell} + t(\Sigma_{jN} + \Sigma_{\ell N}) + t^2 \Sigma_{NN}\,. 
\end{align}
So 	
\be\label{integral_recurrence_1}
\bar{J}^{N}(n;\bm{\nu};\Sigma) = \frac{1}{B(\nu-\nu_{N},\nu_{N}) }
\int_{0}^{\infty}\mathrm{d}t\,\frac{t^{\nu_{N}-1}}{(1+t)^{n-\nu}}
\bar{J}^{N-1}(n - 2\nu_N;\bm{\nu}; \Sigma^{\eta}(t))\,. 
\ee
For example, the 4 point	function in 4 dimensions will be given by the integral of the 3-point function in 2 dimensions. The right hand side contains the vector $\bm{\nu}$ but it is understood that only $N-1$ components are used.

Iteration of Eq.(\ref{R_recurrence_1}) gives 
\begin{align}
 R(a;\bm{b};\bm{z})&=
\int_{0}^{\infty}
\frac{\mathrm{d}t_{k+1}\ldots \mathrm{d}t_{N}}{B(b_1+\ldots+b_k,b_{k+1},\ldots,b_N)}
\left( \prod_{j=k+1}^{N} t_{j}^{b_j-1}\right)	
\left(1 + \sum\limits_{j=k+1}^{N}t_j \right)^{-a'}  \nonumber \\ 
& \,\,\,\,\,\,\,\,\,\,\,\,\,\,R(a;b_1,\ldots,b_k;z_1+\sum\limits_{j=k+1}^{N}t_j z_j,\ldots,z_k+\sum\limits_{j=k+1}^{N}t_j z_j)\,.
\end{align}
If we take $k=1$ we obtain ~\cite[Eq.(7.7)]{Carlson1963}
\be
 R(a;\bm{b};\bm{z})=
\int_{0}^{\infty}
\frac{\mathrm{d}t_{2}\ldots \mathrm{d}t_{N}}{B(b_1,\ldots,b_N)}
\left( \prod_{j=2}^{N} t_{j}^{b_j-1}\right)	
\left(1 + \sum\limits_{j=2}^{N}t_j \right)^{-a'} \left(z_1+\sum\limits_{j=2}^{N}t_j z_j\right)^{-a}\,,	\nonumber
\ee
so
\begin{align}
\bar{J}^{N}(n;\bm{\nu};\Sigma)&= \frac{1}{B(\nu_1,\ldots,\nu_N)}\expectation \left[
\int_{0}^{\infty}
\mathrm{d}t_{2}\ldots \mathrm{d}t_{N}
\left( \prod_{j=2}^{N} t_{j}^{\nu_j-1}\right)	
\frac{\left(1 + \sum\limits_{j=2}^{N}t_j \right)^{\nu -n}}{\left(i Z_1+\sum\limits_{j=2}^{N} i t_j Z_j\right)^{2\nu-n}}\right]\,,	\nonumber \\
&=
\int_{0}^{\infty}
\frac{\mathrm{d}t_{2}\ldots \mathrm{d}t_{N}}{2\Gamma(2\nu-n)B(\nu_1,\ldots,\nu_N)}
\left( \prod_{j=2}^{N} t_{j}^{\nu_j-1}\right) \nonumber \\ 
&\frac{\Gamma(\nu-n/2) \left(1 + \sum\limits_{j=2}^{N}t_j \right)^{\nu-n}}{\left(\Sigma_{11} +2 \sum\limits_{j=2}^{N} t_j\Sigma_{1j} + \sum\limits_{j,k=2}^{N} t_j\Sigma_{jk}t_k\right)^{\nu-n/2}}\,,
\end{align}
where we have used Eq.(\ref{Schwinger_parameter}) to introduce an integration variable $\tau$, taken the expectation using Eq.(\ref{characteristic_function_multinormal}), changed variable to $v=\tau^2$  and finally 
reapplied Eq.(\ref{Schwinger_parameter}) to get the last line. This form is exactly the form used in ~\cite{vanHameren2011} to compute the $N$-point functions numerically.

Another representation for the $R$ function is given in ~\cite[Eq.(7.8)]{Carlson1963}
\be\label{R_carlson_3}
R(a;\bm{b};\bm{z}) = 
\int_{0}^{1}\frac{\mathrm{d}u u^{b_1-1}(1-u)^{b_2-1}}{B(b_1,b_2) } R(a;b_1+b_2,b_3,\ldots,b_N;u z_1 + (1-u)z_2,\ldots,z_N)\,,
\ee
provided that $b_1$ and $b_2$ have positive parts. Iteration of this integral representation gives
\be
 R(a;\bm{b};\bm{z}) = 
\int_{0}^{1} \frac{\mathrm{d}u_1\ldots \mathrm{d}u_k}{B(b_1,b_2,\ldots,b_k)}
\delta\left( \sum\limits_{j=1}^{k}u_j -1\right)
 \prod_{j=1}^{k} u_j^{b_j-1}
R(a;\sum\limits_{j=1}^{k}b_i,b_{k+1}\ldots,b_N;\sum\limits_{j=1}^{k}u_j z_j,\ldots,z_N)\,,
\ee
and when $k=N$
\be
 R(a;\bm{b};\bm{z}) = 
\int_{0}^{1} \frac{\mathrm{d}u_1\ldots \mathrm{d}u_N}{B(b_1,b_2,\ldots,b_N)} 
\delta\left( \sum\limits_{j=1}^{N}u_j -1\right)
 \prod_{j=1}^{N} u_j^{b_j-1}
\frac{1}{(\sum\limits_{j=1}^{N}u_j z_j)^{a}}\,,
\ee
which is the standard form in term of Feynman parameters.
Using Eq.(\ref{R_carlson_3}) we get
\be\label{integral_recurrence_2}
\bar{J}^{N}(n;\bm{\nu};\Sigma) =
\int_{0}^{1} \frac{\mathrm{d}u u^{\nu_1-1}(1-u)^{\nu_2-1}}{B(\nu_1,\nu_2)}
\bar{J}^{N-1}(n;\nu_1, \nu_2, \ldots, \nu_{N-2}, \nu_{N-1} + \nu_{N};\overline{\Sigma}(u))\,. 
\ee
The components of the matrix $\overline{\Sigma}$ are
\begin{align}
\overline{\Sigma}_{ij} &=  \Sigma_{ij}\,, 			&&i\,,j<N-1\,,\,\nonumber \\
\overline{\Sigma}_{N-1,j} &= u\Sigma_{N-1,j} + (1-u) \Sigma_{Nj}\,,							&&j<N-1\,, \nonumber \\
\overline{\Sigma}_{N-1,N-1} &= u^2\Sigma_{N-1,N-1} + (1-u)^2 \Sigma_{NN} +
2u(1-u)\Sigma_{N-1,N}\,.
\end{align}
Because of the term $\Sigma_{N-1,N}$ which can be negative, the new mass squared $\overline{\Sigma}_{N-1,N-1}$ can be negative.

\section{Probabilistic Interpretation}
In Eq.(\ref{J_N_v_integration})	we make the change of variable $x_i = v u_i$, 
\beqna
J^{N}(n;\bm{\nu};\Sigma) &=& \frac{(-1)^{-\nu}}{2^{\nu -\nOverTwo-1}\prod_i \Gamma(\nu_i)}
\int_{0}^{\infty}\mathrm{d}x_1 x_1^{\nu_1-1} \ldots \int_{0}^{\infty}\mathrm{d}x_N x_N^{\nu_N-1}   
\nonumber \\
&&\int_{0}^{\infty}\, \mathrm{d}v\, v^{\nu-n-1} \delta\left( \frac{1}{v}\sum\limits_{i=1}^{N} x_i -1\right) e^{-\half x\T .\Sigma. x }\,.
\eeqna
We can rewrite the delta function as
\be
\delta\left( \frac{1}{v}\sum\limits_{i=1}^{N} x_i -1\right) =
v \delta\left(\sum\limits_{i=1}^{N} x_i -v\right)\,,
\ee
which forces us to set $v =\sum\limits_{i=1}^{N} x_i$ in the integrand. We have
\be\label{JN_proba}
J^{N}(n;\bm{\nu};\Sigma) =
(-1)^{-\nu}\mathcal{N}
\int_{-\infty}^{\infty}\prod_{i=1}^{N}\mathrm{d}x_i\, 
\frac{\prod_{i=1}^{N}\left(x_i^{\nu_i-1} \, \mathbbm{1}_{\{x_i>0\}}\right)}{2^{\nu -\nOverTwo-1}\prod_i \Gamma(\nu_i)} \left( \sum\limits_{i=1}^{N} x_i\right)^{\nu -n}
\, \frac{e^{-\half x\T .R^{-1}.x}}{\mathcal{N}}\,,
\ee
with $R = \Sigma^{-1}$ and $\mathcal{N} = (2\pi)^{\frac{N}{2}}|R|^{\half}$. The exponential in the integrand is the probability density of the multivariate normal distribution, so the $N$-point function in $n$ dimension is given by the product of the truncated moments of correlated normal random variables times the power of their sum.
\be\label{JN_proba2}
J^{N}(n;\bm{\nu};\Sigma) =
\frac{(-1)^{-\nu}\mathcal{N}}{2^{\nu -\nOverTwo-1}\prod_i \Gamma(\nu_i)}
\expectation \left[
\prod_{i=1}^{N}\left(\epsilon_i^{\nu_i-1} \, \mathbbm{1}_{\{\epsilon_i>0\}}\right) \left( \sum\limits_{i=1}^{N} \epsilon_i\right)^{\nu -n} \right]\,,
\ee
where $\epsilon_i, i=1,\ldots,N$ are normal random variables with covariance matrix $R = \Sigma^{-1}$.

We now consider some special cases.

($\bm{\nu_i =1}$ and $\bm{\nu -n  <0.}$) 
\be
J^{N}(n;\bm{\nu};\Sigma) =  \frac{(-1)^{-\nu}\mathcal{N}}{2^{\nu -\nOverTwo-1}\prod_i\Gamma(\nu_i)}
\expectation \left[ \frac{\prod_i \mathbbm{1}_{\epsilon_i>0}}{\left(\sum\limits_i \epsilon_i\right)^{n-\nu}}\right]\,.
\ee
	
($\bm{\nu_i =1}$ and $\bm{\nu =N = n.}$)	
In this case we have
\be\label{JN_nu_equal_n_probabilist} 	
J^{N}(n;\bm{1};\Sigma) =\frac{(-1)^{-N}\mathcal{N}}{2^{\nu -\nOverTwo-1}} 
\mathrm{Prob}\left( \epsilon_1 >0, \ldots, \epsilon_N >0 \right)\,. 
\ee
	
($\bm{\nu_i = 1}$ and $\bm{\nu - n = k >0.}$)
We expand 
\be
\left(\sum\limits_{i=1}^{N} x_i\right)^k = \sum\limits_{\sum k_i=k}\, \frac{k!}{\prod_i k_i!}\prod_i x_i^{k_i}\,,
\ee 
and get
\be
J^{N}(n;\bm{\nu};\Sigma) = \frac{(-1)^{-N}\mathcal{N}}{2^{\nu -\nOverTwo-1}\prod_i\Gamma(\nu_i)}
\sum\limits_{\sum k_i=k}\, \frac{k!}{\prod_i k_i!}
\,\expectation \left[ \prod\limits_{i=1}^{N}\left[\mathbbm{1}_{\{\epsilon_i>0\}} \epsilon_i^{\nu_i+k_i+1}\right]\right]\,. 
\ee

\section{Explicit Evaluation of $N$-Point Functions}
To find explicit expressions we follow the method of ~\cite{Childs1967} where Fourier Transform methods are used. Recall Eq.(\ref{JN_proba}) 	
\be
J^{N}(n;\bm{\nu};\Sigma)=
\frac{(-1)^{-\nu}}{2^{\nu -\nOverTwo-1}\prod_i\Gamma(\nu_i)}
\frac{(2\pi)^{\frac{N}{2}}}{|\Sigma|^{\half}} 
\int_{-\infty}^{\infty}\prod\limits_{i=1}^{N}\mathrm{d} z_i\,
\prod\limits_{i=1}^{N}U(z_i)\,
\frac{\prod_i z_{i}^{\nu_i-1}\left(\sum_i z_i\right)^{\nu-n}}{(2\pi)^{\frac{N}{2}}|R|^{\half}}e^{-\half z^{T}.R^{-1}. z}\,,
\ee
where $U(z_i) = \mathbbm{1}_{z_i>0}$. The $N$ dimensional Fourier Transform $\hat{f}(w)$ of a function $f(x)$ is defined as
\be
\hat{f}(w) = \int_{-\infty}^{\infty}\mathrm{d}^Nx\, f(x) e^{-iw\T .x}\,,
\ee 
and its inverse is
\be
f(x) = \frac{1}{(2\pi)^N}\int_{-\infty}^{\infty}\mathrm{d}^N w\,  \hat{f}(w) e^{+iw\T .x}\,.
\ee 
both Fourier Transform of the unit step(unidimensional) and Gaussian functions (multidimensional)  are known ~\cite{Childs1967,Hansen2014,Tilman2015}
\begin{align}
\hat{U}(w) &= \frac{1}{i} \frac{1}{\w - i\epsilon}\,, \nonumber \\
           &= \frac{1}{i}  \left( i \pi \delta(w) +\mathrm{PV}\frac{1}{w} \right)\,, \nonumber \\
					 &=  \pi \delta(w) +\mathrm{PV}\left(\frac{1}{i w}\right) \,, \\
					 &=  \pi \delta(w) -i \lim_{\epsilon \to 0}\frac{w}{w^2 + \epsilon^2}\,,\nonumber\\
\mathcal{F}\left(
{\frac{1}{(2\pi)^{\frac{N}{2}}|R|^{\half}}e^{-\half z\T .R^{-1}. z}}
\right)(w) &=  e^{-\half w\T .R. w}\,.
\end{align}
In the following, the limit ($\epsilon \to 0$) is assumed even if not written explicitly. Moreover the Fourier Transform satisfies
\be
\mathcal{F}\left[x_k f(x) \right](w) = i\frac{\mathrm{\partial}}{\partial w_k}\hat{f}(w)\,.
\ee 
The Parseval relation states
\be
\int_{\infty}^{\infty}\mathrm{d}^Nx\,  f(x) g(x) = 
\frac{1}{(2\pi)^N}\int_{\infty}^{\infty}\mathrm{d}^N w\,  \hat{f}(w) \hat{g}(-w)\,,  
\ee
so we get
\beqna
J^{N}(n;\bm{\nu};\Sigma)&=&
\frac{(-1)^{-\nu}}{2^{\nu -\nOverTwo-1}\prod_i\Gamma(\nu_i)}
\frac{(2\pi)^{\frac{N}{2}}}{|\Sigma|^{\half}} 
\frac{1}{(2\pi)^{N}}\int_{-\infty}^{\infty}\prod\limits_{j=1}^{N}\mathrm{d} w_j\,
\left[
\prod\limits_{j=1}^{N}\left( \pi \delta(\omega_j) - \frac{i}{\omega_j} \right)
\right. \nonumber \\
&& \label{N-point_fourier} \left.
\prod\limits_{j=1}^{N} \left(i\frac{\partial }{\partial \omega_j}\right)^{\nu_j-1}
\left(i\sum\limits_{j=1}^{N}\frac{\partial }{\partial \omega_j}\right)^{\nu-n}
\right]
e^{-\half w^{T}\Sigma^{-1} w}\,.
\eeqna

We introduce the following notation for the integral of the Gaussian kernel weighted by denominators 
\be
I_{ij\ldots}^{kl\ldots} =  \int_{-\infty}^{\infty}
 \mathrm{d}\omega_i \mathrm{d}\omega_j \mathrm{d}\omega_k \mathrm{d}\omega_l\ldots
\frac{1}{\omega_i \omega_j\ldots} e^{\w\T .R^{ijkl\ldots}. \w}\,,
\ee 
which means that we integrate the $(i,j,k,l)$ $\omega$ variables with $\w_i$ and $\w_j$ appearing in the denominator with $R^{ijkl}$ the inverse of the covariance matrix $\Sigma$.

\subsection{Two-point Function in 2$D$}
The Fourier integral we need to compute is (with $R = \Sigma^{-1}$)
\be
F_2 = \frac{1}{(2\pi)^{2}}\int_{-\infty}^{\infty} \mathrm{d} w_1\mathrm{d} w_2\,
\left( \pi \delta(w_1) +\frac{1}{i w_1} \right)
\left( \pi \delta(w_2) +\frac{1}{i w_2} \right)
e^{-\half \left( R_{11}w_1^2 + 2 R_{12}w_1w_2 + R_{22} w_2^2 \right)}\,.
\ee
The matrix $R$ has components
\begin{align}
R_{11} &= \frac{ m_2^2}{\Delta^{(2)}}\,,\nonumber \\
R_{22} &=  \frac{m_1^2}{\Delta^{(2)}}\,,  \nonumber \\
R_{12} &= - \frac{m_1 m_2 c_{12}}{\Delta^{(2)}}\,,  \nonumber \\
\Delta^{(2)} &= m_1^2 m_2^2 \left( 1- c_{12}^2\right) =  m_1^2 m_2^2 \sin^2 \tau_{12}\,. \nonumber 
\end{align}
Two terms are null because they involve an odd power of $w_i$ and the product of the two delta functions gives a constant
\be
F_{2} = \frac{1}{4}  -\frac{1}{4\pi^2}\int_{-\infty}^{\infty} \mathrm{d} w_1\mathrm{d} w_2\,
\frac{e^{-\half \left( R_{11}w_1^2 + 2 R_{12}w_1w_2 + R_{22} w_2^2 \right)}}{w_1 w_2}\,.
\ee
We have ~\cite[reproduced here in the appendix]{Childs1967}
\be
\int_{-\infty}^{\infty} \mathrm{d} w_1\mathrm{d} w_2\,
\frac{e^{-\half \left( R_{11}w_1^2 + 2 R_{12}w_1w_2 + R_{22} w_2^2 \right)}}{w_1 w_2}
= -2\pi \arcsin\left( \frac{R_{12}}{\sqrt{R_{11}R_{22}}}\right)\,.
\ee
The two-point function is 
\begin{align}
J^{2}(2;\bm{1};\Sigma) &=
\frac{2\pi}{m_1 m_2 \sin \tau_{12}}
\left( \frac{1}{4} +\frac{1}{2\pi}\arcsin\left(\frac{R_{12}}{\sqrt{R_{11}R_{22}}}\right)\right)\,,\\
&=\frac{\tau_{12}}{m_1 m_2 \sin \tau_{12}}\,,
\end{align}
where we have used
\be\label{arcsin}
\arcsin(z) = -\frac{\pi}{2} + \arccos(-z)\,,
\ee
to get the last line, which is exactly the same expression as Eq.(4.3) derived in ~\cite{DavydychevDelbourgo1998} (modulo an $i\pi^{\frac{n}{2}}$ factor that we included in our definition of $J^{N}(n, \bm{\nu}, \Sigma)$).

\subsection{Three-point Function in 3$D$}
The matrix $R$ has components
\begin{align}
R_{11} &= m_2^2m_3^2(1 - c_{23}^2)/ \Delta^{(3)}\,,  \nonumber \\
R_{22} &= m_1^2m_3^2(1 - c_{13}^2)/ \Delta^{(3)}\,,  \nonumber \\
R_{33} &= m_1^2m_2^2(1 - c_{12}^2)/ \Delta^{(3)}\,,  \nonumber \\
R_{12} &= m_1 m_2 m_3^2 \left( c_{13}c_{23} - c_{12}\right)/ \Delta^{(3)}\,, \nonumber \\
R_{13} &= m_1 m_2^2 m_3 \left( c_{12}c_{23} - c_{13}\right)/ \Delta^{(3)}\,, \nonumber \\
R_{23} &= m_1^2 m_2 m_3 \left( c_{12}c_{13} - c_{23}\right)/ \Delta^{(3)}\,, \nonumber \\
\Delta^{(3)}&= m_1^1 m_2^2 m_3^2 \left( 1 - c_{12}^2 - c_{13}^2 - c_{23}^2 +2c_{12}c_{13}c_{23}\right)\,,\nonumber \\
\Delta^{(3)}&= m_1^1 m_2^2 m_3^2 \, D^{(3)}\,.\nonumber 
\end{align}
In Eq.(\ref{N-point_fourier}), 4 terms are null and we get a constant and 3 terms that are exactly like the 2 point  integral
\begin{align}
J^{3}(3;\bm{1};\Sigma) &= \frac{(-1)(2\pi)^{\frac{3}{2}}}{2^{\half}m_1m_2m_3 \sqrt{D^{(3)}}}
\left[ \frac{1}{8} 
+ \frac{1}{4\pi}\arcsin\left(\frac{R_{12}}{\sqrt{R_{11}R_{22}}}\right) \right.\nonumber \\ 
&+ \left.\frac{1}{4\pi}\arcsin\left(\frac{R_{13}}{\sqrt{R_{11}R_{33}}}\right)
+ \frac{1}{4\pi}\arcsin\left(\frac{R_{23}}{\sqrt{R_{22}R_{33}}}\right) 
\right]\,.
\end{align}
Using the relation Eq.(\ref{arcsin}) and the notation in ~\cite{DavydychevDelbourgo1998}
\begin{align}
\cos \Psi_{12}	&=	\frac{R_{12}}{\sqrt{R_{11}R_{22}}} = \frac{c_{12} - c_{13}c_{23}}{\sin\tau_{13} \sin\tau_{23}}\,,\\
\cos\Psi_{13}		&=	\frac{R_{13}}{\sqrt{R_{11}R_{33}}} = \frac{c_{13} - c_{12}c_{23}}{\sin\tau_{12} \sin\tau_{23}}\,,\\
\cos\Psi_{23}		&= 	\frac{R_{23}}{\sqrt{R_{22}R_{33}}} = \frac{c_{23} - c_{12}c_{13}}{\sin\tau_{13} \sin\tau_{12}}\,,\\
\Omega^{(3)}  	&= 	\Psi_{12}+ \Psi_{13}  + \Psi_{23}  -\pi\,,
\end{align}
we get  
\be
J^{3}(3;\bm{1};\Sigma) = -\frac{\pi^{\half}}{2 m_1 m_2 m_3} \frac{\Omega^{(3)}}{\sqrt{\Delta^{(3)}}}\,,
\ee
exactly as Eq.(5.6) in ~\cite{DavydychevDelbourgo1998}.

\subsection{Three-point Function in 2$D$}
In this case in Eq.(\ref{JN_proba}), the term $(\sum\limits_{i=1}^{3}x_i)^{\nu -n}$ contributes to the integral. 
\be
J^{3}(2;\bm{1};\Sigma) = \frac{(-1)(2\pi)^{\frac{3}{2}}}{2m_1m_2m_3 \sqrt{D^{(3)}}}
\frac{1}{(2\pi)^3} \int_{-\infty}^{\infty}\mathrm{d}^3\omega 
\prod\limits_{j=1}^{3}\left[ \pi \delta(i \omega_j) +\frac{1}{\omega_j}\right]
\left( \sum\limits_{j=1}^{3} i\frac{\partial}{\partial\omega_j}\right)e^{\omega^{T}R\omega}\,.
\ee
We expand and keep the non zero terms
\begin{align}
J^{3}(2;\bm{1};\Sigma) &= \frac{(-1)(2\pi)^{\frac{3}{2}}}{2m_1m_2m_3 \sqrt{D^{(3)}}}
\frac{1}{(2\pi)^3}
\left[ 
 (R_{11}+R_{12}+R_{13})( -\pi^2 I_{1} + I_{23}^{1})  \right. \nonumber \\ 
&+  (R_{12}+R_{22}+R_{23})( -\pi^2 I_{2} + I_{13}^{2}) \nonumber \\ 
&+ \left. (R_{13}+R_{23}+R_{33})(- \pi^2 I_{3} + I_{12}^{3} \right]\,,  \nonumber
\end{align}
with
\begin{align}
I_{j} &= 
\frac{\sqrt{2\pi}}{\sqrt{R_{jj}}}\,,\nonumber \\
I_{ij}^{k} &= 
 -\frac{(2\pi)^{\frac{3}{2}}}{\sqrt{R_{kk}}}
\arcsin\left( \frac{\rho_{ij} - \rho_{ik}\rho_{jk}}{\sqrt{1-\rho_{ik}^2}\sqrt{1-\rho_{jk}^2}}\right)\,,\nonumber \\ 
\rho_{ij} &= \frac{R_{ij}}{\sqrt{R_{ii}R_{jj}}}\,. \nonumber
\end{align}
Explicitly
\begin{align}\label{eq:3-point-2D}
J^{3}(2;\bm{1};\Sigma) &= \frac{(2\pi)^{\frac{3}{2}}}{2^{\half}m_1m_2m_3 \sqrt{D^{(3)}}}
\left[ 
 \frac{(R_{11}+R_{12}+R_{13})}{\sqrt{\pi R_{11}}}
\left(\frac{1}{8} + \frac{1}{4\pi} \arcsin\left( \frac{\rho_{23} - \rho_{12}\rho_{13}}{\sqrt{1-\rho_{13}^2}\sqrt{1-\rho_{23}^2}}\right) \right)  \right. \nonumber \\ 
&+  \frac{(R_{12}+R_{22}+R_{23})}{\sqrt{\pi R_{22}}}
\left(\frac{1}{8} + \frac{1}{4\pi} \arcsin\left( \frac{\rho_{13} - \rho_{12}\rho_{23}}{\sqrt{1-\rho_{12}^2}\sqrt{1-\rho_{23}^2}}\right) \right)   \nonumber \\ 
&+ \left. \frac{(R_{13}+R_{23}+R_{33})}{\sqrt{\pi R_{33}}}
\left(\frac{1}{8} + \frac{1}{4\pi} \arcsin\left( \frac{\rho_{12} - \rho_{13}\rho_{23}}{\sqrt{1-\rho_{13}^2}\sqrt{1-\rho_{23}^2}}\right) \right)  \right]\,. 
\end{align}
In ~\cite[section 4.2]{Davydychev2006},  the 3-point function in 2$D$ is also written as a sum of three terms. Each term looks like a 2-point function in 2$D$. The 3-point function in 3$D$ is a linear combination of 2$D$ 2-point functions with coefficients written using the $\tau_{j\ell}$ variables. Eq.(\ref{eq:3-point-2D}) looks very similar, but written explicitly in terms of the $c_{j\ell}$ variables which are directly linked to the kinematical invariants.~\cite{Davydychev2006} also presents results for the 3-point function in 4,5$D$.

\subsection{Four-point Function in 4$D$}
We write
\be
|\Sigma|^{\half} = m_1 m_2 m_3 m_4 \sqrt{D^{(4)}}\,.
\ee
In Eq.(\ref{N-point_fourier}), 8 terms are null and we get a constant and 6 terms that are exactly like the 2 point  integral plus a integral that involves all the variables $\wi$.
we have
\begin{align}
J^{4}(4;\bm{1};\Sigma) &= \frac{2\pi^2}{m_1 m_2 m_3 m_4 \sqrt{D^{(4)}}}
\, \frac{1}{16} \, \left[ 1 
+ \frac{2}{\pi}\arcsin\left(\frac{R_{12}}{\sqrt{R_{11}R_{22}}}\right) \right.\nonumber \\ 
&+ \left.\frac{2}{\pi}\arcsin\left(\frac{R_{13}}{\sqrt{R_{11}R_{33}}}\right)
+ \frac{2}{\pi}\arcsin\left(\frac{R_{14}}{\sqrt{R_{11}R_{44}}}\right)  \right. \nonumber \\
&+ \left.\frac{2}{\pi}\arcsin\left(\frac{R_{23}}{\sqrt{R_{22}R_{33}}}\right)
+ \frac{2}{\pi}\arcsin\left(\frac{R_{24}}{\sqrt{R_{22}R_{44}}}\right) \right. \nonumber \\
&+ \left. \frac{2}{\pi}\arcsin\left(\frac{R_{34}}{\sqrt{R_{33}R_{44}}}\right)  
+ \frac{1}{\pi^4}I_{1234}
\right]\,.
\end{align}
In the appendix it is shown that $I_{1234}$ is the sum of 3 integrals
\begin{align}
\frac{1}{\pi^4} I_{1234} &= \frac{4\rho_{12}}{\pi^2}\int_{0}^{1}\frac{\mathrm{d}u}{\sqrt{1-\rho_{12}^2u^2}}
\,\arcsin\left( \frac{\alpha_{34}(u)}{\sqrt{\alpha_{33}(u)\alpha_{44}(u)}}\right)\nonumber \\
&+
\frac{4\rho_{13}}{\pi^2}\int_{0}^{1}\frac{\mathrm{d}u}{\sqrt{1-\rho_{13}^2u^2}}
\,\arcsin\left( \frac{\beta_{24}(u)}{\sqrt{\beta_{22}(u)\beta_{24}(u)}}\right) \nonumber \\
&+
\frac{4\rho_{14}}{\pi^2}\int_{0}^{1}\frac{\mathrm{d}u}{\sqrt{1-\rho_{14}^2u^2}}
\,\arcsin\left( \frac{\gamma_{23}(u)}{\sqrt{\gamma_{22}(u)\gamma_{33}(u)}}\right)\,,
\end{align}
with
\begin{align}
\alpha_{33} &= (1- \rho_{23}^2) + u^2(2\rho_{12}\rho_{13}\rho_{23} - \rho_{12}^2 - \rho_{13}^2) \,, \nonumber \\
\alpha_{34} &= (\rho_{34}- \rho_{23}\rho_{24}) + u^2(\rho_{12}\rho_{13}\rho_{24} + \rho_{12}\rho_{14}\rho_{23} - \rho_{13}\rho_{14} - \rho_{12}^2\rho_{34}) \,, \nonumber \\
\alpha_{44} &= (1- \rho_{24}^2) + u^2(2\rho_{12}\rho_{14}\rho_{24} - \rho_{12}^2 - \rho_{14}^2) \,, \nonumber
\end{align}
\begin{align}
\beta_{22} &= (1- \rho_{23}^2) + u^2(2\rho_{13}\rho_{12}\rho_{23} - \rho_{13}^2 - \rho_{12}^2) \,, \nonumber \\
\beta_{24} &= (\rho_{24}- \rho_{23}\rho_{34}) + u^2(\rho_{13}\rho_{12}\rho_{34} + \rho_{13}\rho_{14}\rho_{23} - \rho_{12}\rho_{14} - \rho_{13}^2\rho_{24}) \,, \nonumber \\
\beta_{44} &= (1- \rho_{34}^2) + u^2(2\rho_{13}\rho_{14}\rho_{34} - \rho_{13}^2 - \rho_{14}^2) \,, \nonumber
\end{align}
\begin{align}
\gamma_{22} &= (1- \rho_{24}^2) + u^2(2\rho_{14}\rho_{12}\rho_{24} - \rho_{14}^2 - \rho_{12}^2) \,, \nonumber \\
\gamma_{23} &= (\rho_{23}- \rho_{24}\rho_{34}) + u^2(\rho_{14}\rho_{12}\rho_{34} + \rho_{14}\rho_{13}\rho_{24} - \rho_{12}\rho_{13} - \rho_{14}^2\rho_{23}) \,, \nonumber \\
\gamma_{33} &= (1- \rho_{34}^2) + u^2(2\rho_{14}\rho_{13}\rho_{34} - \rho_{14}^2 - \rho_{13}^2) \,, \nonumber
\end{align}
where
\begin{align}
\rho_{ij} &= \frac{R_{ij}}{\sqrt{R_{ii}R_{jj}}}\,,\nonumber\\
&= \frac{\Delta_{ij}}{\sqrt{\Delta_{ii}\Delta_{jj}}}\,.
\end{align}
$\Delta_{ij}$ is the $(i,j)$ cofactor of the matrix $\Sigma$. This formula is not explicitly symmetric with respect to the indices $(1,2,3,4)$ because we have chosen to transform the denominator $\omega_1$. We could have symmetrised the final expression by cyclic permutation of the indices. The symmetrised result represents 12 integrals of arcsine and square root functions whose values in the complex plane are known. In ~\cite{DavydychevDelbourgo1998}, the geometrical interpretation worked easily for the 2- and 3-point functions. For the 4-point function in dimension 4, the computation of the volume of a four dimensional tetrahedron is quite complicated. In general a symmetric result can be written by decomposing the tetrahedron into 12 so called bi-rectangular tetrahedron. The volume of a bi-rectangular tetrahedron can be expressed in terms of Lobachevsky or 
Sch\"afli functions which can be related to dilogarithms. The volume of the 4 dimensional tetrahedron is, according to our probabilistic interpretation, related to the quadrivariate normal orthant probability. This fact has already been exploited in ~\cite{Abrahamson1964}, where the quadrivariate normal orthant probability is computed using the decomposition of the four dimensional tetrahedron into bi-rectangular tetrahedron which are called orthoschemes in ~\cite{Abrahamson1964}. Using Fourier Transforms, allows us to circumvent difficult geometrical decomposition. For $N\geq 5$, the geometrical interpretations seems to become unrealisable, while Fourier Transform still 
performs well. Besides the geometrical approach to the 4-point function,  direct integration of the Feynman parameters in Eq.(\ref{JN_feynman}) as done 
in ~\cite{THOOFT1979,DENNER1991,DennerDittmaier2011} has produced results in term of a varying number of dilogarithms depending on the assumptions on the internal masses and the momenta. Several transformations are applied to the Feynman representation Eq.(\ref{JN_feynman}) to write the final results as a sum of dilogarithms. The Fourier approach distangles the singularity in the Fourier space to produce contributions to the 4-point function that can be recognised as 2-point functions in 2$D$. From a first look at the remaining integrals, it looks difficult to cast them in term of dilogarithms to make contact with the Feynman parameter integration. The presence of the square root and the arcsine function suggests the integrals might be expressible in terms of elliptic functions.~\cite{vanOldenborgh1990,VANOLDENBORGH1992} presents expressions that are also numerically stable. More efforts are required to implement the above formula in a computer program and study its numerical stability.

\subsection{Five-point Function in 5D}
There are $2^5$ terms of which only 16 survive. With 
$$ \mathrm{det}\Sigma = \prod_{i=1}^{5} m_i^2 D^{(5)}\,,$$ 
\be\label{J5_5D}
J^{5}(5;\bm{1};\Sigma) = \frac{2\pi^{\frac{5}{2}}}{\prod\limits_{i=1}^{5}m_i \sqrt{D^{(5)}}}
\, \frac{1}{32} \, \left[ 
1  + \frac{2}{\pi}\sum\limits_{i<j}\arcsin{\rho_{ij}}
+\frac{1}{\pi^4} \sum\limits_{i=1}^{5} I_{12345\setminus i} \right]\,,
\ee
where $I_{12345\setminus i}$ contains four denominators ans is thus computed as $I_{ijkl}$
with a correlation matrix  $R^{(i)}$ obtained from the matrix $R$ by removing column and row $i$
\be
\frac{1}{\pi^4} I_{12345\setminus i}
= \frac{1}{\pi^4}
\int_{-\infty}^{\infty}\prod\limits_{j\neq i}\mathrm{d}\omega_j
\frac{1}{\prod\limits_{j\neq i}\omega_j}
e^{-\half \omega^{T} R^{(i)}\omega}\,.
\ee

\subsection{Six-point Function in 6D}
There are $2^6$ terms of which only 32 survive.
\be
J^{6}(6;\bm{1};\Sigma) = \frac{2\pi^3}{\prod\limits_{i=1}^{6}m_i \sqrt{D^{(6)}}}
\, \frac{1}{64} \, \left[ 
1  + \frac{2}{\pi}\sum\limits_{i<j}\arcsin{\rho_{ij}}
+\frac{1}{\pi^4} \sum\limits_{i<j} I^{4}_{ij}
- \frac{1}{\pi^6}I_{ijklmn} \right]\,,
\ee
with $I^{4}_{pq} = I_{ijklmn \setminus (p,q)}$.
$I^{4}_{pq}$ is computed exactly as $I_{ijkl}$ as it contains 4 $\omega$s in the denominator. 
Regarding $I_{ijklmn}$, we proceed as for $I_{(ijkl)}$ except that in this case we will need two integrations to decrease the number of $\omega$s in the denominator. $I_{ijklmn}$ is computed in appendix \ref{I6}
\be
\frac{1}{\pi^6}I_{ijklmn}(R) = -\frac{2}{\pi^5}
\int_{0}^{1}\mathrm{d}u
\sum\limits_{r \neq i}
\frac{\rho_{ir}}{\sqrt{1-\rho_{ir}^2 u^2}}I_{jklmn\setminus r}(\tilde{\rho}(u))\,,
\ee
with the matrix $\tilde{\rho}$ 
\be
\tilde{\rho}_{rs}(u) =
\rho_{rs} - \rho_{ir} \rho_{is} u^2 +
\frac{1}{1-\rho^2_{ij}u^2}
\left( \rho_{jr} -\rho_{ij}\rho_{ir} u^2\right)
\left( \rho_{js} -\rho_{ij}\rho_{is} u^2\right)\,.
\ee
In ~\cite{DelDuca2011ne, DelDuca2011jm, DelDuca2011wh} 
the hexagonal diagram in dimension 6 was considered with 0,1 and 3 masses. The results were given in terms of polylogarithms. The 3 mass case result is expressed as a sum of 24 terms involving only one basic function, which is a simple linear combination of logarithms, dilogarithms, and trilogarithms. In our case, the result is also compact and derived straightforwardly. The interest in the 6-point function in dimension 6 stems from its relation to the MHV amplitudes in $\mathcal{N}=4$ SYM theory ~\cite{Dixon2011ng}.

\subsection{Five-point Function in 4D}
The term $(\sum\limits_{i=1}^{5}x_i)^{\nu -n} = (\sum\limits_{i=1}^{5}x_i)$ in  Eq.(\ref{JN_proba}),  contributes to the integral. We replace each $x_i$ by $i\frac{\partial }{\partial \omega_i}$
We get
\be
J^{5}(4;\bm{1};\Sigma) = -\frac{2^{\half}\pi^{\frac{5}{2}}}{\prod\limits_{i=1}^{5}m_i \sqrt{D^{(5)}}}
\frac{1}{(2\pi)^{5}}\int_{-\infty}^{\infty}\mathrm{d}^5\omega
\prod\limits_{j=1}^{5}\left( \pi \delta(\omega_j) - \frac{i}{\omega_j} \right) 
\left( i\sum\limits_{j=1}^{5}\frac{\partial}{\partial \omega_j}\right)
e^{-\half \omega^{T} R \omega}\,.
\ee
Each derivative with respect to $\omega_j$ generates 5 terms so in total we have 
$2^5 \times 5 \times 5 = 800$  terms where only 200 survive
\begin{align}
J^{5}(4;\bm{1};\Sigma) &= -\frac{2^{\half}\pi^{\frac{5}{2}}}{\prod\limits_{i=1}^{5}m_i \sqrt{D^{(5)}}}
\frac{1}{32} \left[  
\frac{1}{\pi}\sum\limits_{i=1}^{ 5}\left(  \sum\limits_{j \leq i} R_{ji} + \sum\limits_{j > i}R_{ij}  \right)I_{i}
\right. \nonumber \\
&+ \frac{1}{\pi^3} \sum\limits_{j>i} \sum\limits_{k\neq i,j} I_{ij}^{k}
\left(  \sum\limits_{l \leq k} R_{lk} + \sum\limits_{l > k}R_{kl}  \right) \nonumber \\
&- \left. \frac{1}{\pi^5}\sum\limits_{i=1}^{5} \left(  \sum\limits_{j \leq i} R_{ji} + \sum\limits_{j > i}R_{ij}  \right) I_{12345 \setminus i}
\right]\,.
\end{align}
This last formula can be interpreted as a decomposition of the 5-point function in 4D in terms of 4-point functions in 4D. Reduction formulae 
have been presented in ~\cite{Halpern1963,Melrose1965,Petersson1965,VANNEERVEN1984,BERN1993}. The decomposition of the 5-point function in 4D in terms of a sum of 4-point functions in 4D was first achieved in ~\cite{Halpern1963}.~\cite{Melrose1965} has generalised the approach to N-point ($N \geq 5$) functions and presented detailed decomposition for the 5,6,7-point functions in 4D. It is remarkable that for $N=5$, the Fourier Transform produces rather simple expressions that can also be interpreted as decomposition formulae.~\cite{VANNEERVEN1984,BERN1993} have also presented original decomposition formulae.

\subsection{Six-point Function in 4D}
The term $(\sum\limits_{i=1}^{6}x_i)^{\nu -n} = (\sum\limits_{i=1}^{6}x_i)^{2}$ in  Eq.(\ref{JN_proba}),  contributes to the integral. We write
\begin{align}
(\sum\limits_{i=1}^{6}x_i)^{2}e^{-\half x^{T} .R^{-1}. x} 
&= (\sum\limits_{i=1}^{6} x_i^2 + 2\sum\limits_{i<j} x_i x_j ) e^{-\half x^{T} .\Sigma. x}\,, \nonumber\\
&= -2 (\sum\limits_{i=1}^{6} \frac{\partial}{\partial \Sigma_{ii}} + \sum\limits_{i<j} \frac{\partial}{\partial \Sigma_{ij}} ) e^{-\half x^{T} .\Sigma. x}\,, \nonumber
\end{align}
so that
\begin{align}
J^{6}(4;\bm{1};\Sigma) &= -\frac{2\pi^3}{\prod\limits_{i=1}^{6}m_i \sqrt{D^{(6)}}}
\left(\sum\limits_{i=1}^{6} \frac{\partial}{\partial \Sigma_{ii}} + \sum\limits_{i<j} \frac{\partial}{\partial \Sigma_{ij}} \right)
\frac{1}{(2\pi)^6}
\int_{-\infty}^{\infty}\mathrm{d}^6 w 
\prod\limits_{i=1}^{6}\left( \pi \delta(\omega_i) - \frac{i}{\omega_i} \right)
e^{-\half \omega^{T} .R. \omega}\,, \nonumber \\
&=-\left(\sum\limits_{i=1}^{6} \frac{\partial}{\partial R_{ii}} + \sum\limits_{i<j} \frac{\partial}{\partial R_{ij}} \right) J^{6}(6,\bm{1},\Sigma)\,.
\end{align}
Unlike ~\cite{Petersson1965} which has produced a decomposition formula in the case $N>n$, which the 6-point function in 4D is an example our formula is written in term of the 6-point function in 6D. It is rather compact, but taking the derivatives will likely generate a large number of terms. It is unclear if the double integrals  
appearing in the 6-point function in 6D  can be simplified so that the 6-point function in 6D can be explicitly decomposed in terms of 4-point functions.

\subsection{ $\nu_i \geq 1$ and $\nu > n$}
In the general case where  $\nu_i > 1$, the $x_i$ variables in the integrand of Eq.(\ref{JN_proba}) will contribute a term
$\prod\limits_{i=1}^{N} x_i^{\nu_i}\left(\sum\limits_{i=1}^{N} x_i\right)^{\nu-n}$
which is also dealt with by replacing each $x_i$ by $i\frac{\partial }{\partial \omega_i}$ so that
\begin{align}
J^{N}(n;\bm{\nu};\Sigma) &= -\frac{(-1)^{-\nu}(2\pi)^{\frac{N}{2}}}{2^{\nu-\frac{n}{2}-1}\prod\limits_{i=1}^{N}\Gamma(\nu_i)m_i \sqrt{D^{(N)}}}
\frac{1}{(2\pi)^N}
\int_{-\infty}^{\infty}\mathrm{d}^N w 
\left[
\prod\limits_{j=1}^{N}\left( \pi \delta(\omega_j) - \frac{i}{\omega_j} \right)
\right. \nonumber \\
& \left.
\prod\limits_{j=1}^{N} \left(i\frac{\partial }{\partial \omega_j}\right)^{\nu_j-1}
\left(i\sum\limits_{j=1}^{N}\frac{\partial }{\partial \omega_j}\right)^{\nu-n}
\right]
e^{-\half \omega^{T} R \omega}\,.
\end{align}
A software package is required to handle the proliferation of terms.

\subsection{$\nu-n>0$}
In this case, we can introduce the Inverse Laplace Transform result
\be\label{inverse_laplace_transform_power}
x^q \mathbbm{1}_{x>0} = \int\limits_{c-i\infty}^{c+i\infty}\frac{\mathrm{d}s}{2\pi i}
\frac{\Gamma(q+1)}{s^{q+1}} e^{s x}\,,
\ee 
with $c>0$ and $\Re(q)>-1$.
Plugging this result, with $ q = \frac{\nu-n}{2}$, in Eq.(\ref{JN_proba}), we obtain
\begin{align}
J^{N}(n<\nu;\bm{\nu};\Sigma) &=
\frac{(-1)^{\nu}}{2^{\nu -\nOverTwo-1}\prod_i\Gamma(\nu_i)}
\int\limits_{c-i\infty}^{c+i\infty}\frac{\mathrm{d}s}{2\pi i}
\frac{\Gamma(\frac{\nu-n}{2}+1)}{s^{\frac{\nu-n}{2}+1}} 
\int_{0}^{\infty}\prod_{i=1}^{N}\mathrm{d}x_i\,\prod_i x_i^{\nu_i-1}\, 
\,e^{-\half x\T .\Sigma.x + s \left( \sum\limits_{i=1}^{N}x_i\right)^2}\,, \nonumber \\
&=
\frac{(-1)^{\nu}}{2^{\nu -\nOverTwo-1}\prod_i\Gamma(\nu_i)}
\int\limits_{c-i\infty}^{c+i\infty}\frac{\mathrm{d}s}{2\pi i}
\frac{\Gamma(\frac{\nu-n}{2}+1)}{s^{\frac{\nu-n}{2}+1}} 
\int_{0}^{\infty}\prod_{i=1}^{N}\mathrm{d}x_i\,\prod_i x_i^{\nu_i-1}\, 
\,e^{-\half \sum\limits_{i,j=1}^{N} x_i\left( \Sigma_{ij} -2 s\right)x_j}\,. \nonumber
\end{align}
The diagonal elements become
\be
\Sigma_{ii} - 2 s = m_i^2 - 2s = \tilde{m}_i^2\,,
\ee
and the non diagonal elements become ($j\neq \ell$)
\begin{align}
\Sigma_{j\ell} - 2 s &= m_j m_{\ell}
\frac{m_j^2 + m_{\ell}^2 - k_{jl}^2}{2 m_j m_{\ell}} - 2s\,,\nonumber\\
&=
\tilde{m}_j \tilde{m}_{\ell}
\frac{\tilde{m}_j^2 + \tilde{m}_{\ell}^2 - k_{jl}^2}{2 \tilde{m}_j \tilde{m}_{\ell}}\,, \nonumber
\end{align}
We introduce the matrix $\tilde{\Sigma}$ which is the same as the
 matrix $\Sigma$ but with the mass parameters $m_i^2$
replaced by $\tilde{m}_i^2$. In this case, we see that the N-point function with $\nu>n$ is obtained by performing the Inverse Laplace Transform of the $N$-point function for which $n=\nu$ and the matrix $\tilde{\Sigma}$ 
\be\label{nu_greater_n}
J^{N}(n<\nu;\bm{\nu};\Sigma) =
\int\limits_{c-i\infty}^{c+i\infty}\frac{\mathrm{d}s}{2\pi i}
\frac{\Gamma(\frac{\nu-n}{2}+1)}{s^{\frac{\nu-n}{2}+1}} 
J^{N}(\nu;\bm{\nu};\tilde{\Sigma}(s))\,,
\ee
with $\tilde{m_i}^2 = m_i^2 -s$ after rescaling $s$ to $s/2$.
In the previous subsection, we have seen that some computations contain many terms. Using this Inverse Laplace transform, we can compute the 6-point function in 4 dimension using the 6-point function in 6 dimension but with complex mass squared $m_i^2 - s$. This Inverse Transform also frees us from being careful in the bookkeeping necessary when the number of terms increase as seen above. This result does not depend on the probabilistic interpretation as it can also be easily derived from the Feynman parameter representation Eq.(\ref{JN_feynman}).

\subsection{$\nu_i\geq 1$ and $\nu < n$}
In this case we use Eq.(\ref{Schwinger_parameter})  and doing as above we end up with the following relation
\be\label{nu_smaller_n}
J^{N}(n>\nu;\bm{\nu};\Sigma) =
\frac{1}{\Gamma(\frac{n-\nu}{2})}
\int_{0}^{\infty}\mathrm{d}\tau\,
\tau^{\frac{n-\nu}{2}-1} 
J^{N}(\nu;\bm{\nu};\tilde{\Sigma}(\tau))\,.
\ee
The matrix $\tilde{\Sigma}(\tau)$  is the same as the matrix $\Sigma$ but with the mass parameters $m_i^2$ replaced by $m_i^2 + \tau$.

\subsection{$\nu_i>1, n=\nu$}
Also of interest are diagrams with some propagator power larger than 1 i.e. $\nu_k>1$. Let us consider the case of an $N$-point function in dimension $n= \nu = N-1 +\nu_k$, with the $k$th propagator having power $\nu_k>1$. In Eq.(\ref{JN_proba}), the term $\sum\limits_i x_i$ disappear but remains a term $x_k^{\nu_k-1}$. We use Eq.(\ref{inverse_laplace_transform_power})
\begin{align}
x_k^{\nu_k-1} &= x_k^{2\frac{{\nu_k-1}}{2}}\,, \nonumber \\
              &= \int\limits_{c-i\infty}^{c+i\infty}\frac{\mathrm{d}s}{2\pi i}
\frac{\Gamma(\frac{\nu_k-1}{2}+1)}{s^{\frac{\nu_k-1}{2}+1}} e^{s x_k^2}\,.   
\end{align}
We end up with
\be\label{nuk_greater_1_nu_eq_n}
J^{N}(N-1 + \nu_k;\nu_j = 1 + (\nu_k-1)\delta_{jk};\Sigma) =
(-1)^{-{\nu_k-1}}
\int\limits_{c-i\infty}^{c+i\infty}\frac{\mathrm{d}s}{2\pi i}\,
\frac{\Gamma(\frac{\nu_k-1}{2}+1)}{s^{\frac{\nu_k-1}{2}+1}}
J^{N}(N;\nu_j=1;\Sigma^k)\,.
\ee
The matrix $\Sigma^k$  is the same as the matrix $\Sigma$ but with the mass parameters $m_k^2$ replaced by $m_k^2 - s$ but only in the component $\Sigma_{kk}$. The other components remain unchanged.  That is $\Sigma_{kk}$ is changed into
\be
\Sigma_{kk} = m_k^2 \to \left(\Sigma^k\right)_{kk} = m_k^2 -s\,.
\ee

Another important case is when two propagators (with index e.g. $k$ and $k'$) have their power greater than 1 in dimension $n=\nu = N-2 + 2\nu_k$. We consider the case when the powers are equal i.e. $\nu_k =\nu_{k'}$. In Eq.(\ref{JN_proba}), we have a term $(x_k x_{k'})^{\nu_k-1}$  that we write as
\be
(x_k x_{k'})^{\nu_k-1} = \int\limits_{c-i\infty}^{c+i\infty}\frac{\mathrm{d}s}{2\pi i}
\frac{\Gamma(\nu_k)}{s^{\nu_k}} e^{s x_k x_{k'}}\,.
\ee
We end up with
\be\label{nuk_eq_nukprime_greater_1_nu_eq_n}
J^{N}(N-2 + 2\nu_k; \nu_j = 1 + (\nu_k-1) \delta_{jk} + (\nu_k-1) \delta_{jk'}; \Sigma) =
\frac{(-1)^{-2(\nu_k-1)}}{\left(2^{\frac{\nu_k-1}{2}}\Gamma(\nu_k)\right)^2}
\int\limits_{c-i\infty}^{c+i\infty}\frac{\mathrm{d}s}{2\pi i}\,
\frac{\Gamma(\nu_k)}{s^{\nu_k}}
J^{N}(N;\nu_j=1;\Sigma^{kk'})\,.
\ee
The matrix $\Sigma^{kk'}$  is the same as the matrix $\Sigma$ but with the component $\Sigma_{kk'}$changed as follows
\begin{align}
\left(\Sigma^{kk'}\right)_{kk'} 
&= m_k m_{k'} \frac{m_k^2 + m_{k'}^2 - k^2_{kk'}}{2m_km_{k'}} -2s\,, \nonumber\\
&= m_k m_{k'} \frac{m_k^2 + m_{k'}^2 - ( k^2_{kk'} + 4s)}{2m_km_{k'}}\,,
\end{align}
that is the kinematical invariant $k^2_{kk'}$ is shifted by the quantity $4s$
\be
k^2_{kk'} \to k^2_{kk'} + 4s\,.
\ee

Eq.(\ref{nuk_eq_nukprime_greater_1_nu_eq_n}, \ref{nuk_greater_1_nu_eq_n}) show that $N$-point functions with higher powers (and $n=\nu$) by integrating in the complex plane $N$-point function 
(with $\nu=N$) with modified $\Sigma$ matrix. Another way to proceed, if we would like to avoid integration in the complex plane is to consider that a propagator with power greater than 1, is the same as a product of standard propagators i.e.
\be
\frac{1}{\left((p_i+q)^2 - m_i^2\right)^{\nu_i}} = \prod_{j=1}^{\nu_i}
\frac{1}{(k_j+q)^2 - M_j^2} \,,
\ee     
with $M_j^2 =  m_i^2$ and $k_j = p_i$. This means that we can use Eq.(\ref{JN_proba}) again 
and obtain 
\be\label{nuk_nukprime_greater_1_nu_eq_n}
J^{N}(N-2 + \nu_k + \nu_{k'}; \nu_j = 1 + (\nu_k-1) \delta_{jk} + (\nu_{k'}-1) \delta_{jk'}; \Sigma) =
J^{N-2 + \nu_k+\nu_{k'}}(N-2 + \nu_k + \nu_{k'}; \nu_j =1; \Sigma^{kk'})\,,
\ee
where the matrix $\Sigma^{kk'}$ is obtained from the matrix $\Sigma$ by adding 
$\nu_k + \nu_{k'}-2$ columns. In these columns the component to be added is
$ m_j m_{n} c_{jn}$ for $j=1,\ldots, N-2 + \nu_k + \nu_{k'}$ and $n=N-1, \ldots,N-2 + \nu_k + \nu_{k'}$.

\section{ $\epsilon$ Expansion}
To show that Eq.(\ref{nu_smaller_n}) does not depend on the probabilistic interpretation, we start from the Feynman parametrisation Eq.(\ref{JN_feynman})
\be
J^{N}(n;\bm{\nu};\Sigma) = \frac{1}{(-1)^{\nu}} \frac{\Gamma(\nu - n/2)}{\prod_i \Gamma(\nu_i)}
\int \prod_i \mathrm{d}u_i u_i^{\nu_i-1} \delta\left( \sum\limits_{i=1}^{N} u_i -1\right)
\frac{1}{\left( -\sum\limits_{j<\ell}u_j u_{\ell} k^2_{j\ell} + \sum u_i m_i^2   \right)^{\nu - \frac{n}{2}}}\,,
\ee
and apply the relation
\begin{equation}
\frac{1}{A^{\lambda_1}B^{\lambda_2}} = 
\frac{\Gamma(\lambda_1 + \lambda_2)}{\Gamma(\lambda_1)\Gamma(\lambda_2)} 
\int_{0}^{\infty}\mathrm{d}x 
\frac{x^{\lambda_2-1}}{\left( A + x B \right)^{\lambda_1+\lambda_2}}\,
\end{equation}
with 
\begin{align}
A&= -\sum\limits_{j<\ell}u_j u_{\ell} k^2_{j\ell} + \sum u_i m_i^2\,,   \nonumber \\
\lambda_1 &= \nu - \frac{n}{2} = \nu - \frac{d}{2} + \epsilon > 0\,, \nonumber \\
B &=1\,, \nonumber \\
\lambda_2 &= k -\epsilon >0\,. 
\end{align}
\begin{align}
J^{N}(n;\bm{\nu};\Sigma) &= 
\frac{1}{\Gamma(k-\epsilon)}\int_{0}^{\infty}\mathrm{d}x x^{k-\epsilon-1} 
(-1)^{\nu} \frac{\Gamma(\nu - \frac{d-2k}{2})}{\prod_i \Gamma(\nu_i)}
\int \prod_i \mathrm{d}u_i u_i^{\nu_i-1} \delta\left( \sum\limits_{i=1}^{N} u_i -1\right)\nonumber \\
&\frac{1}{\left( -\sum\limits_{j<\ell}u_j u_{\ell} k^2_{j\ell} + \sum u_i (m_i^2 + x)   \right)^{\nu - \frac{d-2k}{2}}}\,, \nonumber \\
&\label{epsilon_expansion_integral}=\frac{1}{\Gamma(k-\epsilon)}\int_{0}^{\infty}\mathrm{d}x x^{k-\epsilon-1}
J^{N}(d-2k;\bm{\nu};\Sigma(x))\,, 
\end{align}
where $\Sigma(x)$ as before but with squared mass $m_i^2 +x$ for each line in the diagram. We have obtained a way to express an $N$-point function in decimal dimension as a sum of $N$-point functions in integer dimension, with all possible masses from $m_i^2$ to $\infty$. For example with $N=5$ in dimension $n=6-2\epsilon$ we can write
\begin{align}
J^5(6-2\epsilon;\nu_j=1;\Sigma) &= \frac{1}{\Gamma(1-\epsilon)}
\int_{0}^{\infty}\mathrm{d}x x^{-\epsilon}
J^{5}(4;\nu_j=1;\Sigma(x))\,,\nonumber \\
&= \frac{1}{\Gamma(\half-\epsilon)}
\int_{0}^{\infty}\mathrm{d}x x^{\half-\epsilon-1}
J^{5}(5;\nu_j=1;\Sigma(x))\,. 
\end{align} 
We prefer the expression which involves $J^{5}(5;\nu_i=1;\Sigma(x))$ as it is the one easier to compute in this framework and the one reproduced exactly by Eq.(\ref{nu_smaller_n}). The $\epsilon$ expansion is given by
\be
J^5(6-2\epsilon;\nu_i=1;\Sigma) = \frac{2}{\Gamma(\half-\epsilon)}
\sum\limits_{k=0}^{\infty} \frac{(-2\epsilon)^k}{k!}\, 
\int_{0}^{\infty}\mathrm{d}u \ln(u)^k
J^{5}(5;\nu_i=1;\Sigma(u))\,, 
\ee
with $u =\sqrt{x}$. 

\section{Tensor Integrals}

In ~\cite{Davydychev1991}, a formula was derived for the reduction of tensor to scalar integrals.
We take for example the case of a $N$-point function with tensor of rank 2. We have the following decomposition ~\cite{Davydychev1991}
\begin{align}
J^N_{\mu_1\mu_2}(n; \nu_j;\Sigma)  &= 
−\frac{1}{2} g_{\mu_1 \mu_2} J^N(n+2; \nu_j;\Sigma) \nonumber \\
&+\sum\limits_{k=1}^{N} \, p_{k_{\mu_1}} p_{k_{\mu_2}} \nu_k(\nu_k+1) 
J^N(n+4; \nu_j + 2\delta_{jk};\Sigma) \nonumber \\
& \sum\limits_{k<k'}^{N} \, \left(p_{k_{\mu_1}} p_{k'_{\mu_2}} + p_{k_{\mu_1}} p_{k'_{\mu_2}}\right)
\nu_k \nu_{k'} J^N(n+4; \nu_j + \delta_{jk} + \delta_{jk'};\Sigma)\,.
\end{align}
If we apply this relation for $N=5$ with $n = 4-2\epsilon$, we need to compute three diagrams i.e.
\begin{align}
& J^5(6-2\epsilon; \nu_j=1;\Sigma)\,, \nonumber \\ 
& J^5(8-2\epsilon; \nu_j=1 + 2\delta_{jk};\Sigma)\,, \nonumber \\
& J^5(8-2\epsilon; \nu_j=1 + \delta_{jk} + \delta_{jk'};\Sigma)\,.
\end{align}
Using Eq.(\ref{epsilon_expansion_integral}), we have 
\begin{align}
J^5(6-2\epsilon; \nu_j=1;\Sigma) &= \frac{1}{\Gamma(\half-\epsilon)}\int_{0}^{\infty}\mathrm{d}x x^{\half-\epsilon-1} J^{5}(5;\nu_j=1;\Sigma(x)) \,,  \\ 
J^5(8-2\epsilon; \nu_j=1 + 2\delta_{jk};\Sigma) &= \frac{1}{\Gamma(\half-\epsilon)}\int_{0}^{\infty}\mathrm{d}x \, x^{\half-\epsilon-1} J^{5}(7; \nu_j=1 + 2\delta_{jk};\Sigma(x))\,, \\
J^5(8-2\epsilon; \nu_j=1 + \delta_{jk} + \delta_{jk'};\Sigma) &= \frac{1}{\Gamma(\half-\epsilon)}\int_{0}^{\infty}\mathrm{d}x \, x^{\half-\epsilon-1} J^{5}(7;\nu_j=1 + \delta_{jk} + \delta_{jk'};\Sigma(x))\,,
\end{align}
where we have  used $k=\half$. $J^{5}(5;\nu_j=1;\Sigma(x))$ is computed using 
Eq.(\ref{J5_5D}), $J^{5}(7;\nu_j=1 + 2\delta_{jk};\Sigma(x))$ using Eq.(\ref{nuk_greater_1_nu_eq_n}) or Eq.(\ref{nuk_nukprime_greater_1_nu_eq_n}) and $J^{5}(7;\nu_j=1 + \delta_{jk} + \delta_{jk'};\Sigma(x))$ using Eq.(\ref{nuk_eq_nukprime_greater_1_nu_eq_n}) or Eq.(\ref{nuk_nukprime_greater_1_nu_eq_n})\begin{align}
J^{5}(7;\nu_j=1 + 2\delta_{jk};\Sigma(x)) &= J^{7}(7; \nu_j=1; \left[\Sigma(x)\right]^{kk})\,, \\
J^{5}(7;\nu_j=1 + \delta_{jk} + \delta_{jk'};\Sigma(x)) &= J^{7}(7; \nu_j=1; \left[\Sigma(x)\right]^{kk'})\,,
\end{align}
or
\begin{align}
J^{5}(7; \nu_j=1 + 2\delta_{jk}; \Sigma(x)) &= 
\int\limits_{c-i\infty}^{c+i\infty} \frac{\mathrm{d}s}{2\pi i}\, 
\frac{1}{s^{2}} J^{5}(5;\nu_j=1;\left[\Sigma(x)\right]^{kk}(s))\,, \\
J^{5}(7; \nu_j=1 + \delta_{jk} + \delta_{jk'};\Sigma(x)) &= 
\half  \int\limits_{c-i\infty}^{c+i\infty}\frac{\mathrm{d}s}{2\pi i}\,
\frac{1}{s^2} J^{5}(5; \nu_j=1; \left[\Sigma(x)\right]^{kk'}(s))\,.
\end{align}
By applying two transformations on the matrix $\Sigma$, a $N$-point function with higher powers of the propagators and in decimal dimension is reduced to the computation of an $N'$-point function in integer dimension. Moreover, the $\epsilon$ expansion is explicit and easy to derive. In the case $N=5$, we end up computing a complex integral of a 5-point function or the 7-point function 
$J^{7}(7; \nu_j=1; \Sigma$) which is given by
\begin{align}
J^{7}(7; \nu_j=1; \Sigma) &= 
\frac{-(2\pi)^{\frac{7}{2}}}{2^{\frac{5}{2}}\prod\limits_i m_i\,\sqrt{D^{7}}}
\frac{1}{(2\pi)^7}\int\limits_{-\infty}^{\infty}\mathrm{d}^7\w\, 
\prod\limits_{j=1}^{7} \left( \pi\delta(\wj) -\frac{i}{\wj}\right)
e^{-\half x\T.R.x}\,, \nonumber \\
&= 
\frac{-(\pi)^{\frac{7}{2}}}{2^6\prod\limits_i m_i\,\sqrt{D^{7}}}
\left( 1 + \frac{2}{\pi}\sum\limits_{r,s} \arcsin(\rho_{rs})
+ \frac{1}{\pi^4} \sum\limits_{r,s,t,n} I_{rstn}
+ \frac{1}{\pi^4} \sum\limits_{r} I_{ijklmnp\setminus r}
\right)\,.
\end{align}
This expression involves only quantities we know how to compute.

\section{Conclusion}
Recasting the standard Feynman parameter expression for the $N$-point function into a probability problem allowed us to find new integral recurrence relations between $N$-point functions. We have also derived a Hermite polynomial expansion as a general result. However, the multivariate polynomials are still a challenge for numerical evaluation. The series might still be of interest 
for asymptotic expansions and we hope to spend more efforts on the  analysis of the series. We have also derived a multi-fold integral representation for the $N$-point function that admits a probabilistic interpretation. The $N$-point function is related to the truncated moments of the multivariate normal distribution. This distribution has been extensively studied and many numerical and mathematical methods have been developed for its computation. Other methods could be borrowed from the statistical literature to compute the $N$-point function. Using Fourier Transforms, it was possible to compute several $N$-point functions in integer dimension. The extension to decimal dimension was made possible by introducing an extra variable. The $N$-point function in decimal dimension is given by integrating $N$-point functions in integer dimension but with mass parameters depending on the integration variable. The case of tensor integrals of rank $r=2$, with $N=5$ was treated explicitly. A reduction program was achieved in this case. We leave it to the future to extend the reduction program to tensor of rank $r>2$, with $N>5$.

\section*{Acknowledgements}
I would like to thank Mama Benhaddou for encouragements. 

\renewcommand \thesection{\Alph{section}}
\renewcommand{\theequation}{\Alph{section}.\arabic{equation}}
\setcounter{section}{0}
\setcounter{equation}{0}

\section{Computation of $I_{ij}$}
We define
\be
\rho_{ij} = \frac{R_{ij}}{\sqrt{R_{ii}}\sqrt{R_{jj}}}\,.
\ee
$I_{ij}(R)$ is given by
\be
I_{ij}(R) =\int_{-\infty}^{\infty}\mathrm{d}^2\w
\frac{1}{\w_i \w_j }  \, e^{-\half \sum\limits_{m,n} \w_m R_{mn} \w_n}\,.
\ee
We change variable $w_m \rightarrow \frac{w_m}{\sqrt{R_{mm}}}$ 
\be
I_{ij}(R) =
\int_{-\infty}^{\infty}\mathrm{d}^2\w
\frac{1}{\w_i \w_j }  \, e^{-\half \left(\w_i^2 + \w_j^2  + 2 \rho_{ij} \w_i \w_j \right)}\,.
\ee
For a function $f(\rho)$ we can write
\be
f(\rho) = f(0)  + \int_{0}^{\rho}f'(u)\mathrm{d}u\,,
\ee
where in our case the function $f(\rho)$ is $I_{ij}$ as a function 
of $\rho_{ij}$. For $\rho_{ij} =0$,  $I_{ij}$ is zero, so we are left with
\begin{align}
I_{ij}(R) &= -\int_{-\infty}^{\infty}\mathrm{d}^2\w
\int_{0}^{\rho_{ij}}\mathrm{d}u
\, e^{-\half(\w_i^2 +\w_j^2  + 2u\w_i\w_j)}\,, \\
&=
-2\pi\int_{0}^{\rho_{ij}}\mathrm{d}u
\frac{1}{\sqrt{1-u^2}}\,,\\
&= -2\pi\arcsin\rho_{ij}\,, \\
&= -2\pi\arcsin\frac{R_{ij}}{\sqrt{R_{ii}R_{jj}}}\,. 
\end{align}

\setcounter{equation}{0}
\section{Computation of $I_{ijkl}$}
$I_{ijkl}(R)$ is given by
\be
\frac{1}{\pi^4}I_{ijkl}(R) =\frac{1}{\pi^4}\int_{-\infty}^{\infty}\mathrm{d}^4\w
\frac{1}{\w_i \w_j \w_k \w_l}  \,
e^{-\half \sum\limits_{m,n}\w_m R_{mn} \w_n}\,.
\ee
We rescale the $\w$ variable so that the matrix $R$ has unit numbers in the diagonal
$\w_i = \frac{\sqrt{2} \w_i}{\sqrt{R_{ii}}}$ and use
\be \label{tau_integral}
\frac{1}{\w_i} = \frac{\w_i}{\w_i^2} =  \w_i\int_{0}^{\infty}\mathrm{d}\tau e^{-\tau \w_i^2}\,,  
\ee
which combined  with the already existing $\w_i^2$ in the exponential gives
\be
\frac{1}{\pi^4}I_{ijkl}(R) =\frac{1}{\pi^4}
\int_{1}^{\infty} \mathrm{d}\tau
\int_{-\infty}^{\infty}\mathrm{d}^4\w
\frac{\w_i}{\w_j \w_k \w_l}  \,
e^{-\tau\w_i^2 - \sum\limits_{m\neq i}\w_m^2 - 2\sum\limits_{m<n}\w_m \rho_{mn} \w_n}\,,
\ee
with  (no assumed summation of repeated indices)
\begin{align}
\rho_{mn} &= \frac{R_{mn}}{\sqrt{R_{mm}R_{nn}}}\,,\nonumber\\
\label{rho_matrix}&= \frac{\Delta_{mn}}{\sqrt{\Delta_{mm}\Delta_{nn}}}\,,
\end{align}
where $\Delta_{mn}$ is the $(m,n)$ cofactor of the matrix $\Sigma$. 

We set the functions
\begin{align}
f(\wi) &= \wi e^{-\tau \wi^2}\,, \nonumber \\
g(\wi) &= e^{-2 \wi \left( \rho_{ij}\wj + \rho_{ik}\wk + \rho_{il}\wl\right)}\,, \nonumber
\end{align}
and perform an integration by part whose first contribution is zero and the remaining integral is
\begin{align}
\frac{1}{\pi^4}I_{ijkl}(R) &=-\frac{1}{\pi^4}
\int_{1}^{\infty} \frac{\mathrm{d}\tau}{\tau}
\int_{-\infty}^{\infty}\mathrm{d}^4\w
\left( \frac{\rho_{ij}}{\w_k\w_l} + \frac{\rho_{ik}}{\w_j\w_l} + \frac{\rho_{il}}{\w_j\w_k} \right)
e^{-\tau\w_i^2 - \sum\limits_{m\neq i}\w_m^2 - 2\sum\limits_{m<n} \w_m \rho_{mn} \w_n}\,, \nonumber \\
&=-\frac{1}{\pi^4} 
\int_{1}^{\infty} \frac{\mathrm{d}\tau}{\tau}
\left( \rho_{ij} F_{kl}^{ij}(\tau) + \rho_{ik} F_{jl}^{ik}(\tau) 
+ \rho_{il} F_{jk}^{il}(\tau) \right)\,,\nonumber
\end{align}
 with
\be
F_{kl}^{ij}(\tau) = \int_{-\infty}^{\infty}\mathrm{d}^4\w
\frac{1}{\w_k \w_l}
e^{-\tau\w_i^2 - \sum\limits_{m\neq i}\w_m^2 - 2\sum\limits_{m<n}\w_m \rho_{mn} \w_n}\,.
\ee
We integrate first the $\w$ variables that don't appear in the denominator and get
\begin{align}
F_{kl}^{ij}(\tau) &= 
\int\limits_{-\infty}^{\infty}\mathrm{d}\w_l
\int\limits_{-\infty}^{\infty}\mathrm{d}\w_k
\frac{e^{-\w_k^2 -\w_l^2 -2\rho_{kl}\w_{k} \w_{l} }}{\w_{k} \w_{l}}
\int\limits_{-\infty}^{\infty}\mathrm{d}\w_j
e^{-\w_j^2 -2\w_{j}\left(  \rho_{jk}\w_k + \rho_{jl}\w_l\right)}
\nonumber \\
&\int\limits_{-\infty}^{\infty}\mathrm{d}\w_i
e^{-\tau \w_i^2 -2\w_{i}\left(  \rho_{ij}\w_j +\rho_{ik}\w_k + \rho_{il}\w_l\right)}\,.
\end{align}
With
\be
\int_{-\infty}^{\infty}\mathrm{d}\w  e^{-a \w^2 + b w}
= \frac{\sqrt{\pi}}{\sqrt{a}}e^{\frac{b^2}{4a}}\,,
\ee
we obtain
\begin{align}
F_{kl}^{ij}(\tau) &= 
\int\limits_{-\infty}^{\infty}\mathrm{d}\w_l
\int\limits_{-\infty}^{\infty}\mathrm{d}\w_k
\frac{e^{-\w_k^2 -\w_l^2 -2\rho_{kl}\w_{k} \w_{l} }}{\w_{k} \w_{l}}
\int\limits_{-\infty}^{\infty}\mathrm{d}\w_j
e^{-\w_j^2 -2\w_{j}\left(  \rho_{jk}\w_k + \rho_{jl}\w_l\right)}
\frac{\sqrt{\pi}}{\sqrt{\tau}}e^{\frac{\left(\rho_{ij}\w_j +\rho_{ik}\w_k + \rho_{il}\w_l\right)^2}{\tau}} \nonumber \\
&=\frac{\sqrt{\pi}}{\sqrt{\tau}}
\int\limits_{-\infty}^{\infty}\mathrm{d}\w_l
\int\limits_{-\infty}^{\infty}\mathrm{d}\w_k
\frac{e^{-\w_k^2 -\w_l^2 -2\rho_{kl}\w_{k} \w_{l} + \frac{\left(\rho_{ik}\w_k + \rho_{il}\w_l\right)^2}{\tau} }}{\w_{k} \w_{l} }\,\nonumber \\
&\int\limits_{-\infty}^{\infty}\mathrm{d}\w_j
e^{-\w_j^2(1-\frac{\rho_{ij}^{2}}{\tau}) -2\frac{\w_{j}}{\tau}\left(  \w_k(\tau \rho_{jk} - \rho_{ij}\rho_{ik}) + \w_l(\tau \rho_{jl} - \rho_{ij}\rho_{il})\right)}  \nonumber \\
&=\frac{\pi}{\sqrt{\tau-\rho_{ij}^{2}}}
\int\limits_{-\infty}^{\infty}\mathrm{d}\w_l
\int\limits_{-\infty}^{\infty}\mathrm{d}\w_k
\frac{e^{-\w_k^2 -\w_l^2 -2\rho_{kl}\w_{k} \w_{l} + \frac{\left(\rho_{ik}\w_k + \rho_{il}\w_l\right)^2}{\tau} }}{\w_{k} \w_{l} }e^{\frac{\left(\w_k(\tau \rho_{jk} - \rho_{ij}\rho_{ik}) + \w_l(\tau \rho_{jl} - \rho_{ij}\rho_{il})\right)^2}{\tau(\tau-\rho_{ij}^2)}} \nonumber \\
&=\frac{\pi}{\sqrt{\tau - \rho_{ij}^2}}
\int\limits_{-\infty}^{\infty}\mathrm{d}\w_l
\int\limits_{-\infty}^{\infty}\mathrm{d}\w_k
\frac{1}{\w_k \w_l} e^{-\frac{\alpha_k(\tau) \w_k^2 + \alpha_l(\tau) \w_l^2 + 2\alpha_{kl}(\tau)\w_k\w_l  }{\tau - \rho_{ij}^2}}\,, \nonumber \\
&=
\frac{-2\pi^2}{\sqrt{\tau - \rho_{ij}^2}} 
\arcsin{\frac{\alpha_{kl}(\tau)}{\sqrt{\alpha_{k}(\tau)\alpha_{l}(\tau)}}}\,,
\end{align}
with
\begin{align}
\alpha_{k}(\tau) &= \tau\left(1- \rho_{jk}^2\right) + 2\rho_{ij}\rho_{ik}\rho_{jk} - \rho_{ij}^2 - \rho_{ik}^2 \,, \nonumber \\
&= \alpha_{k}^{0} \tau + \alpha_{k}^{1}\,, \nonumber \\
\alpha_{kl} &= \tau \left(\rho_{kl}- \rho_{jk}\rho_{jl}\right)  + \rho_{ij}\rho_{ik}\rho_{jl} + \rho_{ij}\rho_{il}\rho_{jk} - \rho_{ik}\rho_{il} - \rho_{ij}^2\rho_{kl}\,, \nonumber \\
&= \alpha_{kl}^{0} \tau + \alpha_{kl}^{1}\,, \nonumber \\
\alpha_{l} &= \tau \left(1- \rho_{jl}^2\right)  + 2\rho_{ij}\rho_{il}\rho_{jl} - \rho_{ij}^2 - \rho_{il}^2) \,, \nonumber \\
&= \alpha_{l}^{0} \tau + \alpha_{l}^{1}\,. \nonumber
\end{align}
Bringing all together we get
\begin{align}
\frac{1}{\pi^4}I_{ijkl}(R) &=\frac{4}{\pi^2}
\int_{0}^{1} \mathrm{d}u 
\left( 
\frac{\rho_{ij}}{\sqrt{1-\rho_{ij}^2 u^2}} 
\arcsin{\frac{\alpha_{kl}^{0} + \alpha_{kl}^{1}u^2}{\sqrt{(\alpha_{k}^{0} + \alpha_{k}^{1}u^2)(\alpha_{l}^{0} + \alpha_{l}^{1}u^2)}}} \right. \nonumber \\
&+
\frac{\rho_{ik}}{\sqrt{1-\rho_{ik}^2 u^2}} 
\arcsin{\frac{\alpha_{jl}^{0} + \alpha_{jl}^{1}u^2}{\sqrt{(\alpha_{j}^{0} + \alpha_{j}^{1}u^2)(\alpha_{l}^{0} + \alpha_{l}^{1}u^2)}}} \nonumber \\
\label{I4_final}&+\left.
\frac{\rho_{il}}{\sqrt{1-\rho_{il}^2 u^2}} 
\arcsin{\frac{\alpha_{jk}^{0} + \alpha_{jk}^{1}u^2}{\sqrt{(\alpha_{j}^{0} + \alpha_{j}^{1}u^2)(\alpha_{k}^{0} + \alpha_{k}^{1}u^2)}}} \right)\,,
\end{align}
 where we have made the change of variable $\tau = 1/u^2$.

\setcounter{equation}{0}
\section{Computation of $I_{ijklmn}$}\label{I6}
$I_{ijklmn}(R)$ is given by
\be
\frac{1}{\pi^6}I_{ijklmn}(R) =\frac{1}{\pi^6}\int_{-\infty}^{\infty}\mathrm{d}^6\w
\frac{1}{\w_i \w_j \w_k \w_l \w_m \w_n}  \,
e^{-\half \sum\limits_{r,s}\w_r R_{rs} \w_s}\,.
\ee
We proceed as for $I_{ijkl}(R)$ by rescaling the $\w$ variables which make us use the matrix $\rho$ of Eq.(\ref{rho_matrix}), use Eq.(\ref{tau_integral})
and get (after integration by part as above)
\begin{align}
\frac{1}{\pi^6}I_{ijklmn}(R) &= -\frac{1}{\pi^6}
\int_{1}^{\infty}\frac{\mathrm{d}\tau}{\tau}
\int_{-\infty}^{\infty}\mathrm{d}^6\w
\left(
\frac{\rho_{ij}}{\w_k \w_l \w_m \w_n}
+\frac{\rho_{ik}}{\w_j \w_l \w_m \w_n}
+\frac{\rho_{il}}{\w_j \w_k \w_m \w_n} \right. \\ \nonumber 
&\phantom{-\frac{1}{\pi^6} \int_{1}^{\infty}\frac{\mathrm{d}\tau}{\tau} \int_{-\infty}^{\infty}\mathrm{d}^6\w} 
\left. +\frac{\rho_{im}}{\w_j \w_k \w_l \w_n} +\frac{\rho_{in}}{\w_j \w_k \w_l \w_m}\right) 
e^{-\tau\w_i^2 -\sum\limits_{r\neq i}\w_r^2  -2  \sum\limits_{r<s}\w_r \rho_{rs} \w_s}\,, \nonumber \\
&=-\frac{1}{\pi^6}
\int_{1}^{\infty}\frac{\mathrm{d}\tau}{\tau}
\int_{-\infty}^{\infty}\mathrm{d}^6\w
\left(\sum\limits_{r\neq i} \frac{\rho_{ir}}{\prod\limits_{s \neq (i,r)}\w_s}
\right) e^{-\tau\w_i^2 -\sum\limits_{r\neq i}\w_r^2  -2  \sum\limits_{r<s}\w_r \rho_{rs} \w_s}\,, \nonumber\\
&=-\frac{1}{\pi^6}
\int_{1}^{\infty}\frac{\mathrm{d}\tau}{\tau}
\sum\limits_{r\neq i} \rho_{ir} H_{jklmn\setminus r}^{ir}(\tau)\,, 
\end{align}
with $jklmn\setminus r $ means we remove index that is 
equal to $r$. Taking $r=j$,
\begin{align}
H_{klmn}^{ij}(\tau) &=
\int_{-\infty}^{\infty}\mathrm{d}^4\w
\frac{e^{-\sum\limits_{r,s} \wr \rho^{(i,j)}_{rs}\ws}}{\wk\wl\wm\wn}\,\int_{-\infty}^{\infty}\mathrm{d}\wj
e^{-\wj^2 -2\sum\limits_{q}\wj \rho_{jq}\wq} \int_{-\infty}^{\infty}\mathrm{d}\wi
e^{-\tau\wi^2 -2\sum\limits_{p}\wi \rho_{ip}\wp}\,,
\end{align}
where $\rho^{(i,j)}$ is the matrix $\rho$ with column and row $(i,j)$ removed so that it is a $(4,4)$ matrix for the variables $(k,l,m,n)$. Because the indices $(i,j)$ are always different from the indices $(r,s)$, we have 
$\rho^{(i,j)}_{rs} = \rho_{rs}$. The indices $(q,r,s$) are in the set $(k,l,m,n)$ and the index $p$ goes over $(j,k,l,m,n)$. 
The variables $\w_i$ and $\w_j$ do not appear in the denominator and are integrated first
\begin{align}
H_{klmn}^{ij}(\tau) &=
\int_{-\infty}^{\infty}\mathrm{d}^4\w
\frac{e^{-\sum\limits_{r,s} \wr \rho^{(i,j)}_{rs}\ws}}{\wk\wl\wm\wn}
\int_{-\infty}^{\infty}\mathrm{d}\wj
e^{-\wj^2 -2\sum\limits_{q}\wj \rho_{jq}\wq}
\frac{\sqrt{\pi}}{\sqrt{\tau}} e^{\frac{\left( \rho_{ij}\wj + \sum\limits_{q} \rho_{iq}\wq \right)^2}{\tau}}\,, \nonumber \\
&=\int_{-\infty}^{\infty}\mathrm{d}^4\w
\frac{e^{-\sum\limits_{r,s} \wr \rho^{(i,j)}_{rs}\ws 
+\frac{1}{\tau}\left(\sum\limits_{q}\rho_{iq}\wq\right)^2 }}
{\wk\wl\wm\wn}\nonumber \\
&\frac{\sqrt{\pi}}{\sqrt{\tau}}
\int_{-\infty}^{\infty}\mathrm{d}\wj
e^{-\wj^2\left(1-\frac{\rho_{ij}^2}{\tau}\right) 
-2\frac{\wj}{\tau}\left(\sum\limits_{q} 
\left(\tau \rho_{jq} - \rho_{ij}\rho_{iq} \right)\wq\right)}\,, \nonumber \\
&=\int_{-\infty}^{\infty}\mathrm{d}^4\w
\frac{e^{-\sum\limits_{r,s} \wr \rho^{(i,j)}_{rs}\ws 
+\frac{1}{\tau}\left(\sum\limits_{q}\rho_{iq}\wq\right)^2 }}
{\wk\wl\wm\wn}
\frac{\pi}{\sqrt{\tau}\sqrt{1-\frac{\rho_{ij}^2}{\tau}}}
e^{\frac{\left(\sum\limits_{q} \wq\left(\tau \rho_{jq} - \rho_{ij}\rho_{iq} \right)\right)^2}{\tau(\tau-\rho^2_{ij})}}\,, \nonumber \\
&=
\frac{\pi}{\sqrt{\tau}\sqrt{1-\frac{\rho_{ij}^2}{\tau}}}
\int_{-\infty}^{\infty}\mathrm{d}^4\w
\frac{e^{-\sum\limits_{r,s} \wr \rho^{(i,j)}_{rs}\ws 
+\frac{1}{\tau}\sum\limits_{r,s}\wr \rho_{ir}\rho_{is}\ws}}
{\wk\wl\wm\wn}
e^{\frac{
\sum\limits_{r,s} \wr
\left(\tau \rho_{jr} - \rho_{ij}\rho_ir \right)
\left(\tau \rho_{js} - \rho_{ij}\rho_is \right) \ws
}{\tau(\tau-\rho^2_{ij})}}\,, \nonumber \\
& \nonumber \\
&=
\frac{\pi}{\sqrt{\tau}\sqrt{1-\frac{\rho_{ij}^2}{\tau}}}
\int_{-\infty}^{\infty}\mathrm{d}^4\w
\frac{e^{-\sum\limits_{r,s} \wr \tilde{\rho}_{rs}\ws }}{\wk\wl\wm\wn}\,,
\end{align}
where the matrix $\tilde{\rho}$ has components ($\tau = \frac{1}{u^2}$)
\be
\tilde{\rho}_{rs}(u) =
\rho_{rs} - \rho_{ir} \rho_{is} u^2 +
\frac{1}{1-\rho^2_{ij}u^2}
\left( \rho_{jr} -\rho_{ij}\rho_{ir} u^2\right)
\left( \rho_{js} -\rho_{ij}\rho_{is} u^2\right)\,.
\ee
The final result is
\be
\frac{1}{\pi^6}I_{ijklmn}(R) = -\frac{2}{\pi^5}
\int_{0}^{1}\mathrm{d}u
\sum\limits_{r \neq i}
\frac{\rho_{ir}}{\sqrt{1-\rho_{ir}^2 u^2}}I_{jklmn\setminus r}(\tilde{\rho}(u))\,.
\ee
$ I_{jklmn\setminus r}\left( \tilde{\rho} \right)$ is an integral with 4 denominators and a $4\times 4$ matrix $\tilde{\rho}$. We compute $I_{jklmn\setminus r}(\tilde{\rho}(u))$ using Eq.(\ref{I4_final}). $I_{ijklmn}(R)$ is thus given by 15 double integrals with integrand containing square roots and  $\arcsin$ functions whose analytical continuation in the complex plane is known.



\bibliography{bibliography}
\bibliographystyle{JHEP}

\end{document}